\documentclass[usenatbib]{mn2e}
\usepackage{graphicx}
\usepackage{natbib}
\usepackage{amssymb}
\usepackage{aas_macros}

\def\degg{\hbox{$\null^\circ$\hskip-3pt.}}

\title[Exploring the Andies] {Structural properties of the M31 dwarf
  spheroidal galaxies}\author [McConnachie
\& Irwin] {A.  W.  McConnachie$^{1,2}$
  \& M.  J.  Irwin$^1$\\
  $^1$Institute of Astronomy, Madingley Road, Cambridge, CB3 0HA, U.K.\\
  $^2$Department of Physics and Astronomy, University of Victoria,
  Victoria, BC V8P 5C2, Canada}

\begin{document}

\maketitle

\begin{abstract}
  
  The projected structures and integrated properties of the
  Andromeda~I, II, III, V, VI, VII and Cetus dwarf spheroidal galaxies
  are analysed based upon resolved counts of red giant branch stars.
  The observations were taken as part of the Isaac Newton Telescope
  Wide Field Survey of M31 and its environs.  For each object, we have
  derived isopleth maps, surface brightness profiles,
  intensity-weighted centres, position angles, ellipticities, tidal
  radii, core radii, concentration parameters, exponential scale
  lengths, Plummer scale lengths, half-light radii, absolute
  magnitudes and central surface brightnesses. Our analysis probes to
  larger radius and fainter surface brightnesses than most previous
  studies and as a result we find that the galaxies are generally
  larger and brighter than has previously been recognised. In
  particular, the luminosity of Andromeda~V is found to be consistent
  with the higher metallicity value which has been derived for it.  We
  find that exponential and Plummer profiles  provide
  adequate fits to the surface brightness profiles, although the more 
  general King models provide the best formal
  fits.  Andromeda~I shows strong evidence of tidal disruption and
  S-shaped tidal tails are clearly visible.  On the other hand, Cetus
  does not show any evidence of tidal truncation, let alone
  disruption, which is perhaps unsurprising given its isolated
  location.  Andromeda~II shows compelling evidence of a large excess
  of stars at small radius and suggests that this galaxy consists of a
  secondary core component, in analogy with recent results for
  Sculptor and Sextans.  Comparing the M31 dwarf spheroidal population
  with the Galactic population, we find that the scale radii of the
  M31 population are larger than those for the Galactic population by
  at least a factor of two, for all absolute magnitudes. This
  difference is either due to environmental factors or orbital
  properties, suggesting that the ensemble average tidal field
  experienced by the M31 dwarf spheroidals is weaker than experienced
  by the Galactic dwarf spheroidals. We find that the two populations
  are offset from one another in the central surface brightness --
  luminosity relation, which is probably related to this difference in
  their scale sizes. Finally, we find that the M31 dwarf spheroidals
  show the same correlation with distance-from-host as shown by the
  Galactic population, such that dwarf spheroidals with a higher
  central surface brightness are found further from their host. This
  again suggests that environment plays a significant role in dwarf
  galaxy evolution, and requires detailed modelling to explain the
  origin of this result.

\end{abstract}


\section{Introduction}

Most of our detailed knowledge on the structure of dwarf spheroidal
(dSph) galaxies comes from observations of the
nine\footnote{\cite{willman2005} have recently discovered a new
  Galactic companion in Ursa Major, which is not included here} dSphs
which make up part of the Galactic satellite system.  \cite{irwin1995}
(hereafter IH95) analysed the structure of the Galactic dSphs, except
Sagittarius, by mapping their resolved star counts from photographic
plates. They found that the stellar profiles of the dSphs were
generally well described by a single component King or exponential
model. A generic feature was an excess of stars at large radii, over
and above that expected from the best-fit King tidal model, which have
generally been interpreted as evidence for tidal disturbance. 
A King profile is not a physically motivated model for dSphs since their
relaxation time is of order a Hubble time and their stellar velocity 
distribution may deviate significantly from Maxwellian. Any interpretation of
the structure of a dSph based exclusively on a King model fit must
therefore be treated with caution. Nevertheless, such analyses do
yield a useful parameterisation, and in particular the equivalent half-light
radius can readily be compared with alternative
parameterisations derived from Plummer or exponential profiles.
The King model tidal radii ($r_t$) of the Galactic dSphs
are of order 1\,kpc, and their average half-light radius
($r_{\frac{1}{2}}$) is $\sim 170$\,pc, although Fornax is over twice
this size. More recent work on the radial profiles of Galactic dSphs
from wide field CCD cameras (Ursa Minor:
\citealt{majewski2000,martinezdelgado2001b}; Draco:
\citealt{odenkirchen2001}; Carina: \citealt{palma2003}) generally
confirm the results from earlier photographic studies.

Some recent studies have revealed that the structures of dSphs are more
complex than had previously been assumed. \cite{harbeck2001} looked
for spatial variations in the colour of the RGB and the horizontal
branch of several Local Group dSphs and found evidence for horizontal
branch population gradients in many of their
systems. \cite{tolstoy2004} have shown that Sculptor has a spatially,
chemically and kinematically distinct second component, in the form of
a centrally concentrated core. Additionally, \cite{kleyna2004} have
shown that the Sextans dSph has a kinematically cold core, and that
Ursa Minor has a distinct concentration of stars offset from its
geometric centre which has distinct kinematics
(IH95; \citealt{kleyna2003}).

Given these recent findings in the Galactic dSph system, it is timely
to look at the next closest dSph system to our own, that of M31.
Approximately 16 satellites make up this subsystem, of which seven -
Andromeda~I, II, III, V, VI, VII, and IX - are classified as dSphs.
Andromeda~I, II and III were discovered by Sidney van den Bergh in his
pioneering survey for Local Group galaxies in the early 1970's
(\citealt{vandenbergh1972a,vandenbergh1972b,vandenbergh1974}),
together with Andromeda~IV, an object which was later shown to be a
background galaxy (\citealt{ferguson2000}). Andromeda~V and VI were
discovered some years later by \cite{armandroff1998,armandroff1999}
from a detailed analysis of the digitised Second Palomar Sky Survey.
At the same time, Andromeda~VI was discovered independently by
\cite{karachentsev1999} along with with Andromeda~VII. An additional 
object, Andromeda~VIII (\citealt{morrison2003}), has recently been 
postulated to be associated
with an over-density of planetary nebulae in the giant stellar stream
visible in the south-east of M31 (\citealt{ibata2001a}), although the
nature of this over-density remains unclear. Andromeda~IX was recently
discovered by \cite{zucker2004a} and is the faintest satellite of M31
yet known, with $M_v \simeq -8.3$.

In comparison to the dSphs which orbit the Galaxy, relatively little
is known in detail about the stellar populations of the Andromeda
dSphs. From the colour of their red giant branches, it would appear
that all of them are metal poor [Fe/H]\,$\lesssim$\,-1.5\,dex
(eg. \citealt{mcconnachie2005a}). Deep HST fields 
for Andromeda~I, II and III reaching below the horizontal branch
(\citealt{dacosta1996,dacosta2000,dacosta2002}) show extended epochs of star
formation and variations in horizontal branch morphology, suggesting
the star formation histories have been notably different. In
addition, Andromeda~I is observed to display a gradient in its
horizontal branch morphology, such that their are more blue horizontal
branch stars located at larger radius from the centre of the dwarf
(\citealt{dacosta1996}). Some evidence for AGB components have also
been seen in these systems (most recently by \citealt{kerschbaum2004}
and \citealt{harbeck2004}), although a strong intermediate age
component similar to some of the Galactic dSphs is generally lacking.

As a compliment to the deeper pointed HST observations, we have obtained
wide field Johnston V ($V^\prime$) and Gunn i ($i^\prime$) photometry
for the majority of the members of the M31 subgroup using the Wide
Field Camera (WFC) on the 2.5m Isaac Newton Telescope (INT). This is a
four-chip EEV 4K x 2K CCD mosaic camera with a $\sim 0.29
\square^\circ$ field of view (\citealt{walton2001}). With typical
exposures of $\sim 1000$\,s in each passband, this photometry is deep
enough to observe the top few magnitudes of the red giant branch in
each system and has already been used to derive a homogeneous set of
distance and metallicity estimates for each of these galaxies
(\citealt{mcconnachie2004a,mcconnachie2005a}). Here, we use the same
data to analyse the structural properties of each of the Andromeda
dSphs, as well as the isolated dSph in Cetus as a comparison. In
common with Tucana, this galaxy is one of only two dSphs not found as
a satellite to a large galaxy in the Local Group. Our results for
Andromeda~IX are presented elsewhere (\citealt{chapman2005}). Our
overall technique and methodology is similar to that adopted by IH95
for the Galactic dSphs, insofar as we base most of our analysis on resolved
star counts. \cite{caldwell1992} and \cite{caldwell1999} have
performed a similar analysis for the Andromeda dSphs based upon the
integrated light, and we will compare our results to these later
(Section~4.1).

In Section~2, we derive contour maps, radial profiles and associated
structural parameters for Andromeda I, II, III, V, VI, VII and Cetus
based upon resolved star counts. The integrated luminosities, central
surface brightnesses and related quantities are derived in
Section~3. We postpone discussion of all the results until Section~4,
which also compares the M31 dSph population to the Galactic dSph
population. Section~5 summarises and concludes.

\section{Star counts and structure}
\label{counts}

\subsection{Preliminaries}

The reader is referred to \cite{ferguson2002} and
\cite{mcconnachie2004a,mcconnachie2005a} for information on our
observing strategy and data reduction process. We do, however, recap
here the morphological photometric classification procedure for each
detected source, as this is central to our subsequent analysis.

~

{\bf\noindent Object Classification:} Objects are classified
independently in each passband based on their overall morphological
properties, specifically their ellipticity as derived from
intensity-weighted second moments and the curve-of-growth of their
flux distribution (\citealt{irwin2004}).  Measures from these are
combined to produce a normalised N(0,1) statistic which compares the
likeness of each object to the well-defined stellar locus visible on
each frame. Stellar objects are generally chosen to lie within 2 or
$3\,\sigma$ of this locus depending on the desired trade-off between
completeness and contamination from non-stellar objects. Contamination
takes the form of a small number of spurious images which are
essentially eliminated by requiring objects to be detected on both
passbands. Additionally, at faint magnitudes (within \mbox{$\simeq 1 -
  2$\,mags} of the frame limit depending on seeing), distant compact
galaxies may also be incorrectly identified as stellar images. This is
particularly true in bluer passbands. For this specific study,
completeness is more important than mild non-stellar contamination.
We therefore use all objects identified in the INT~WFC that lie within
$3\,\sigma$ of the stellar locus in the $i^\prime$-band, and require
only that the object has been detected in the $V^\prime$-band within
1\,arcsec of the $i^\prime$-band position, to reduce contamination
from background galaxies.

~

\begin{figure}
\begin{center}
\includegraphics[angle=0, width=7.5cm]{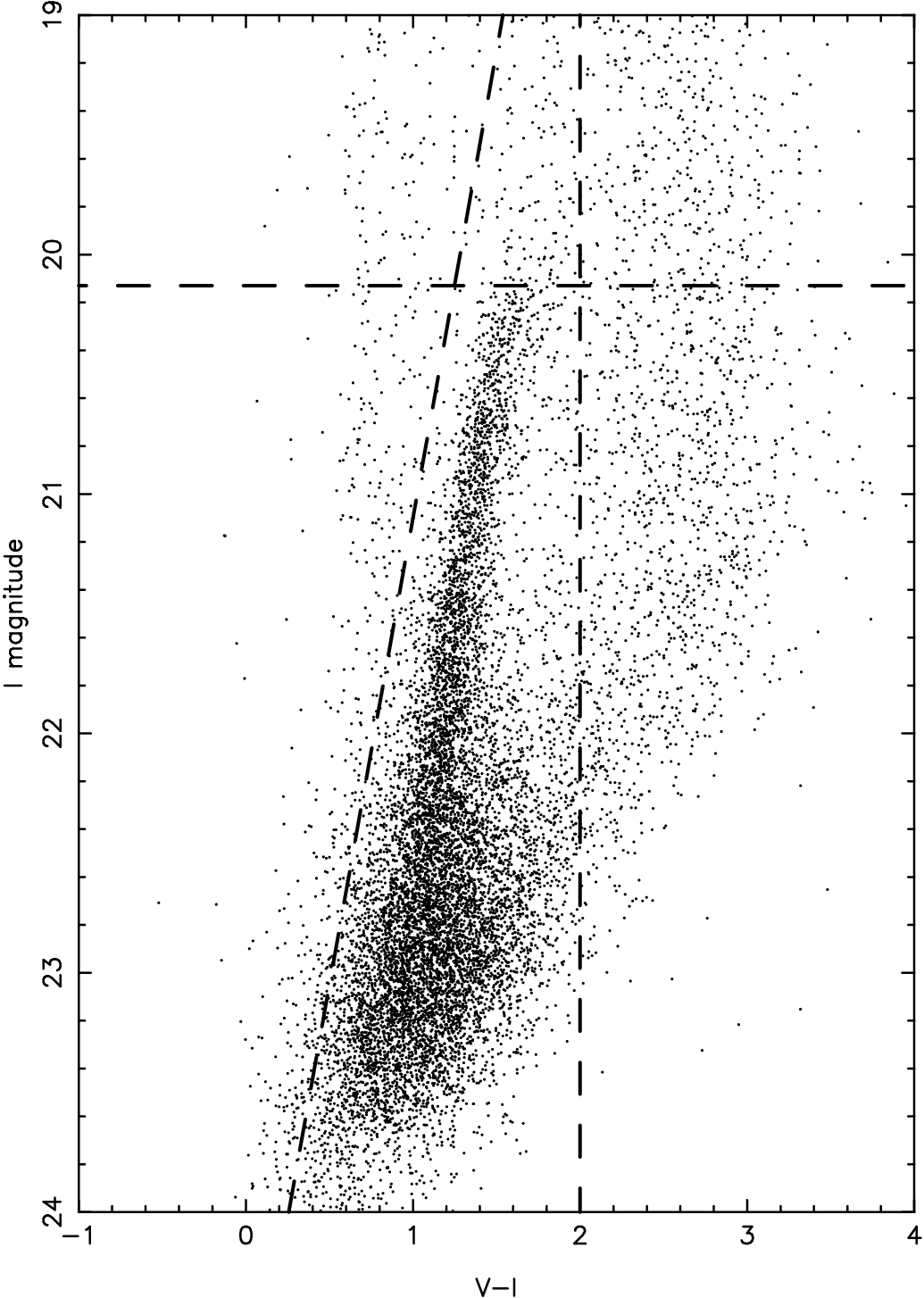}
\caption{A colour magnitude diagram of Andromeda II in Landolt $V$ and
$I$. The dashed lines represent the cuts that we use in order to
reduce background contamination. The horizontal cut is at the location
of the tip of the red giant branch derived in McConnachie et
al. (2004, 2005). The other two cuts are placed so as to isolate the
red giant branch loci as shown.}
\label{andii}
\end{center}
\end{figure}

{\bf\noindent Background corrections:} A good estimate of the
background contamination and its uncertainty is vital to accurately
determining the extent and profile of each dSph. The term
``background'' describes both Galactic foreground stars and distant
compact galaxies. These are minimised in the first instance by using
simple cuts in colour -- magnitude space, designed to isolate the RGB
and remove those stars whose colour and magnitude make membership of
the dSph unlikely. This is illustrated for Andromeda~II in
Figure~\ref{andii}. A horizontal cut in $i^\prime$ magnitude,
coincident with the derived location of the TRGB, removes objects too
bright to be RGB stars in the dSph. Two cuts on the blue and red side
of the RGB also removes objects whose colour makes membership of the
dSph unlikely. This process removes a significant fraction of the
Galactic foreground.

Although the influence of compact galaxies is significantly reduced by
the morphological classification/selection scheme, it cannot be
completely removed at faint magnitudes. A statistical correction must
therefore be applied to remove the remaining contamination. This is
estimated by excising the dSph from a pixelated map of the stellar
spatial distribution. An intensity distribution is constructed from
the remaining pixels and a sigma-clipped least-squares fit of a
Gaussian profile is performed on this ``background'' distribution to calculate
the position of the peak (ie. the background level) and associated
errors in the usual way. As we do not know the full extent of the dSph
{\it a priori}, the correct subtraction of the dwarf component
requires several attempts before we are left with the easily
identifiable background component to perform the fit to. We do not
require to simultaneously fit the dwarf profile plus background
because the background is relatively flat over the area of the field,
with no evidence of strong differential extinction for any of the
dwarfs which would affect this measurement.  For large dSphs which
cover the majority of the field of view (Andromeda~II, VII and Cetus), the
background estimate will suffer from contamination from the outer
regions of these objects. This effect is generally small, as the number
density of stars belonging to the dSph at large radius is much smaller than
the background contamination.  However, this effect will lead to a slight 
overestimate of the background by a few percent, resulting in a marginal 
steepening of the outer parts of
the derived profile. The level of this effect is small enough to be
covered by the derived uncertainties on the scale-radii, and will
always act to underestimate the actual values of these parameters.

~

{\bf\noindent Crowding corrections:} In the central regions of the
dSphs, the higher density of images causes some individual sources to
remain unresolved.  This can be approximately corrected for by using 
the crowding correction of \cite{irwin1984},

\begin{equation}
f^\prime \simeq f \left( 1 + 2 f A^\prime +\frac{16}{3} f^2 A^{\prime 2} ... \right)
\label{trimble}
\end{equation}

\noindent where $f$ is the observed number density of all images (not
just stellar), $f^\prime$ is the actual number density and $A^\prime$
is the typical area of the image, where the radius of the image is
closely approximated by the seeing. For the INT~WFC observations, the
typical seeing is $\simeq 1$\,arcsec. The derivation and validity of
Equation~\ref{trimble} is given in the Appendix of \cite{irwin1984}
and ignores second-order effects such as the actual shape of the
luminosity function. For the dSphs, the typical background stellar
densities are $\sim$ 2 -- 3 stars\,arcmin$^{-2}$ (after photometric
cuts have been applied) and are unchanged by the crowding
correction. The crowding correction increases the stellar counts in
the central regions of the dSphs by typically $5 - 10\,\%$.

\subsection{Isopleths and dwarf geometry}
\label{isopleths}

A field of $0\degg5 \times 0\degg5$ containing the dSph is used for
each isopleth map. For Andromeda~II, V, VI, VII and Cetus, this
corresponds closely to the INT~WFC pointing, and the north-west corner
is missing in each case due to the geometry of the CCDs. For
Andromeda~I and III, multiple pointings of the camera are combined to
form a mosaic.  For each object, the field is divided into $\sim 100 \times
100$\,pixels ($\simeq 18$\,arcsecs resolution), and the crowding
correction is applied based upon the source density in each cell. The
isopleth maps of the stellar sources are constructed, smoothed and
displayed in Figure~\ref{maps}. For the two fields constructed from
multiple INT~WFC pointings, faint-end threshold magnitudes have been
used to account for varying incompleteness levels between
fields due to different observing conditions.  Contour levels in 
Figure~\ref{maps} are set such that each
subsequent increment is a factor $n$ greater than the previous
increment, where the first increment is $2\,\sigma$ above the
background. $n$ is different for each system and is given in the
caption of Figure~\ref{maps}. This allows the same number of contours
to be used for each dwarf galaxy, making visual comparison of their
morphologies easier. The $2\,\sigma$ contour also picks out the peaks of
noise features as well as the dSph, and gives a general
indication of the quality of the background subtraction.

For each isophote, we calculate the centre of gravity $(x_o,\,y_o)$,
position angle, $\theta$ (measured east from north), and eccentricity,
$e$, by using the intensity weighted moments. Thus,

\begin{eqnarray}
x_o&=&\frac{\sum_i x_i I\left(x_i,\,y_i\right)}{\sum_i I\left(x_i,\,y_i\right)}~~~~~y_o~=~\frac{\sum_i y_i I\left(x_i,\,y_i\right)}{\sum_i I\left(x_i,\,y_i\right)}\nonumber\\
~\nonumber\\
\theta&=&\frac{1}{2}\arctan\frac{2 \sigma_{xy}}{\sigma_{yy} - \sigma_{xx}}\nonumber\\
~\nonumber\\
e&=&\frac{\sqrt{\left(\sigma_{xx} - \sigma_{yy}\right)^2 + 4 \sigma^2_{xy}}}{\sigma_{xx} + \sigma_{yy}}
\end{eqnarray}

\noindent where $I\left(x_i,\,y_i\right)$ is the intensity of the
$i^{th}$ pixel and $\sigma_{xx}$, $\sigma_{yy}$ and $\sigma_{xy}$ are
the intensity-weighted second moments within each isophotes. The
eccentricity is related to the ellipticity, $\epsilon = 1 - b/a$, via

\begin{equation}
\epsilon = 1 - \sqrt{\frac{1 - e}{1 + e}}~.
\end{equation}

\noindent The ``average'' parameters for each dSph are then estimated
from the values derived for each isophote and the uncertainty
estimated from the variation in the parameters as a function of
isophotal threshold. This procedure gives estimates of $\theta$ and
$\epsilon$ that are independent of any parameterisation, and allows a
more robust estimate of their uncertainties. Effects such as isophote
rotation are likewise able to be quantified for those systems where
such an effect is taking place. The results of this analysis are
tabulated in Table~\ref{geom}.

\begin{table*}
\begin{center}
\begin{tabular*}{0.65\textwidth}{rrcccc c }
\hline
    &&& \multicolumn{3}{c}{Shape Parameters} \\
               && RA ($x_o$)                                & Dec ($y_o$)              & PA $= \theta$ (\,$^\circ$)&&  $\epsilon = 1 - b/a$\\
\hline
&&&&\\
 Andromeda I   && $00{\rm h} 45{\rm m} 40.3{\rm s}$ & $+38^\circ 02' 21''$ &  22  $\pm$  15 && 0.22  $\pm$  0.04 \\
&&&&\\
 Andromeda II  && $01{\rm h} 16{\rm m} 28.3{\rm s}$ & $+33^\circ 25' 42''$ &  34  $\pm$   6 && 0.20  $\pm$  0.08 \\
&&&&\\
 Andromeda III && $00{\rm h} 35{\rm m} 31.1{\rm s}$ & $+36^\circ 30' 07''$ & 136  $\pm$   3 && 0.52  $\pm$  0.02 \\
&&&&\\
 Andromeda V   && $01{\rm h} 10{\rm m} 17.0{\rm s}$ & $+47^\circ 37' 46''$ &  32  $\pm$  10 && 0.18  $\pm$  0.05 \\
&&&&\\
 Andromeda VI  && $23{\rm h} 51{\rm m} 46.9{\rm s}$ & $+24^\circ 34' 57''$ & 163  $\pm$   3 && 0.41  $\pm$  0.03 \\
&&&&\\
 Andromeda VII && $23{\rm h} 26{\rm m} 33.5{\rm s}$ & $+50^\circ 40' 48''$ &  94  $\pm$   8 && 0.13  $\pm$  0.04 \\
&&&&\\
 Cetus         && $00{\rm h} 26{\rm m} 10.5{\rm s}$ & $-11^\circ 02' 32''$ &  63  $\pm$   3 && 0.33  $\pm$  0.06 \\
&&&&\\
\hline
\end{tabular*}
\caption{Intensity-weighted centres and geometric
  parameters for the dSphs. Errors are estimated from the variation in
  the parameters as a function of isophotal threshold. Centres are
  estimated to be accurate to within $\pm 7$\,arcsecs in both
  directions.}
\label{geom}
\end{center}
\end{table*}

\begin{figure*}
\begin{center}
\includegraphics[angle=270, width=7.cm]{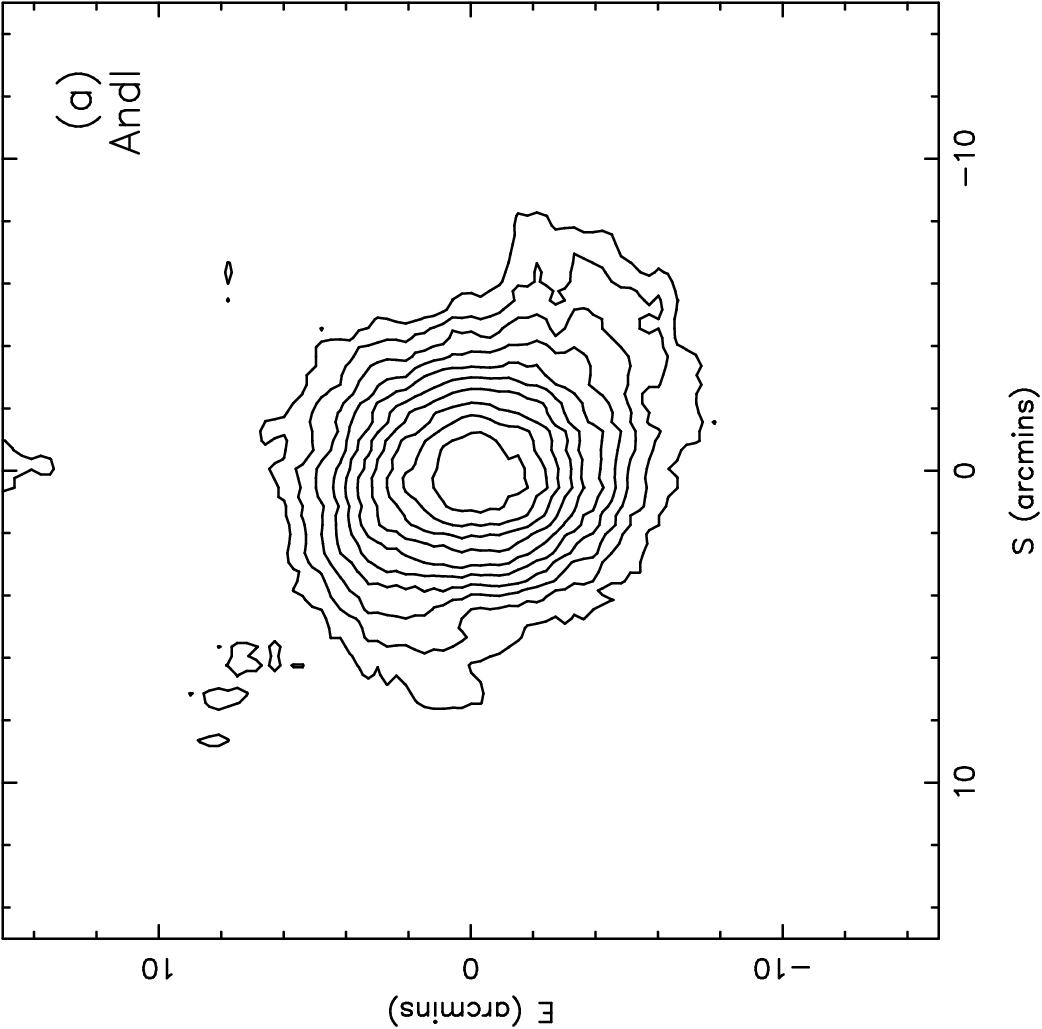}\hspace{5mm}
\includegraphics[angle=270, width=7.cm]{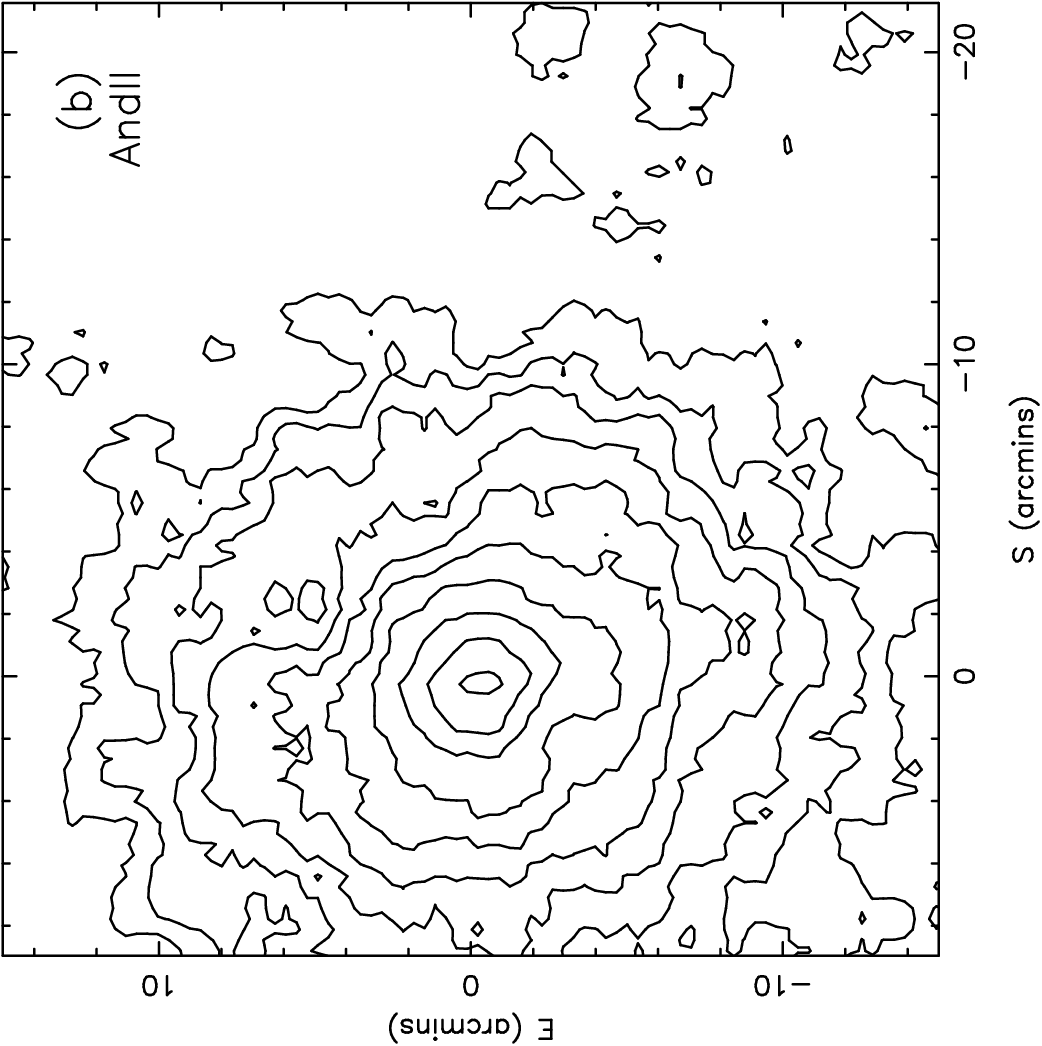}\vspace{5mm}
\includegraphics[angle=270, width=7.cm]{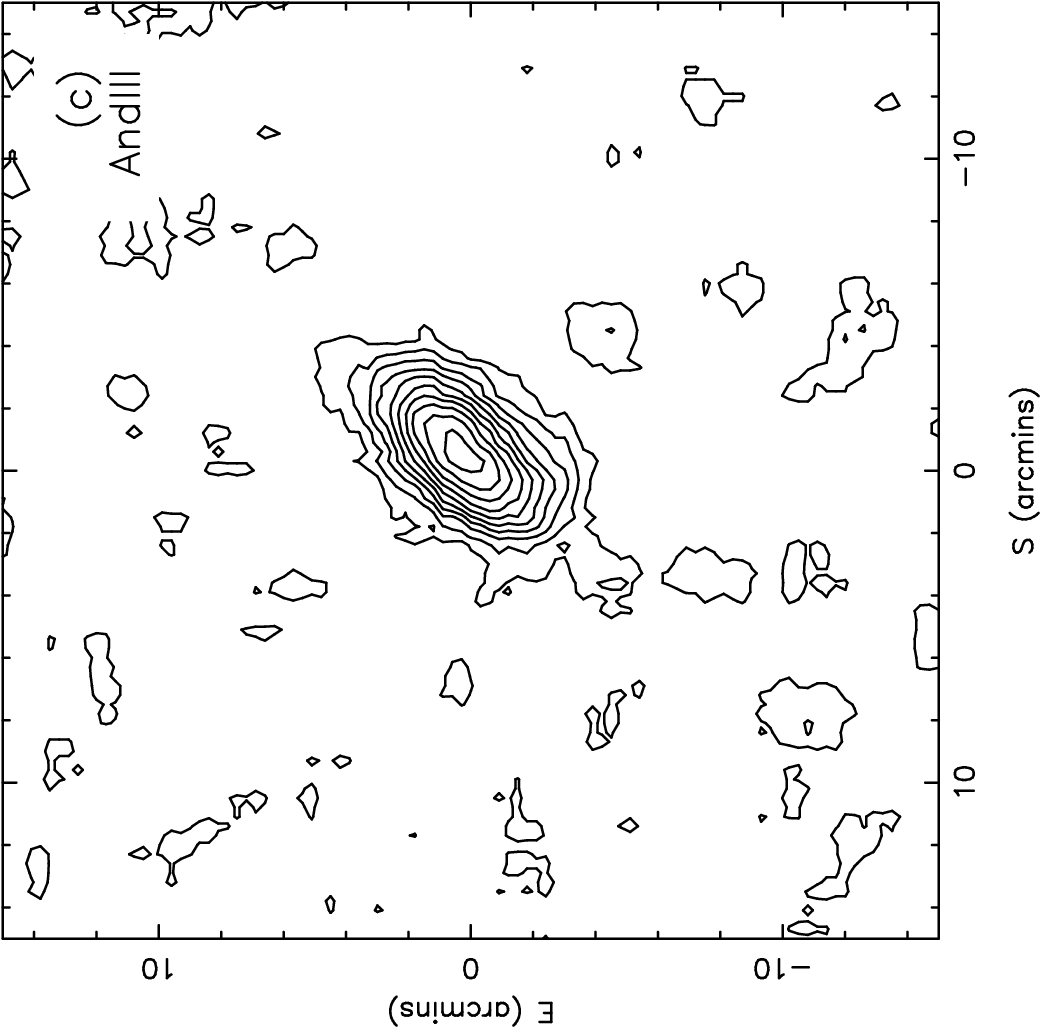}\hspace{5mm}
\includegraphics[angle=270, width=7.cm]{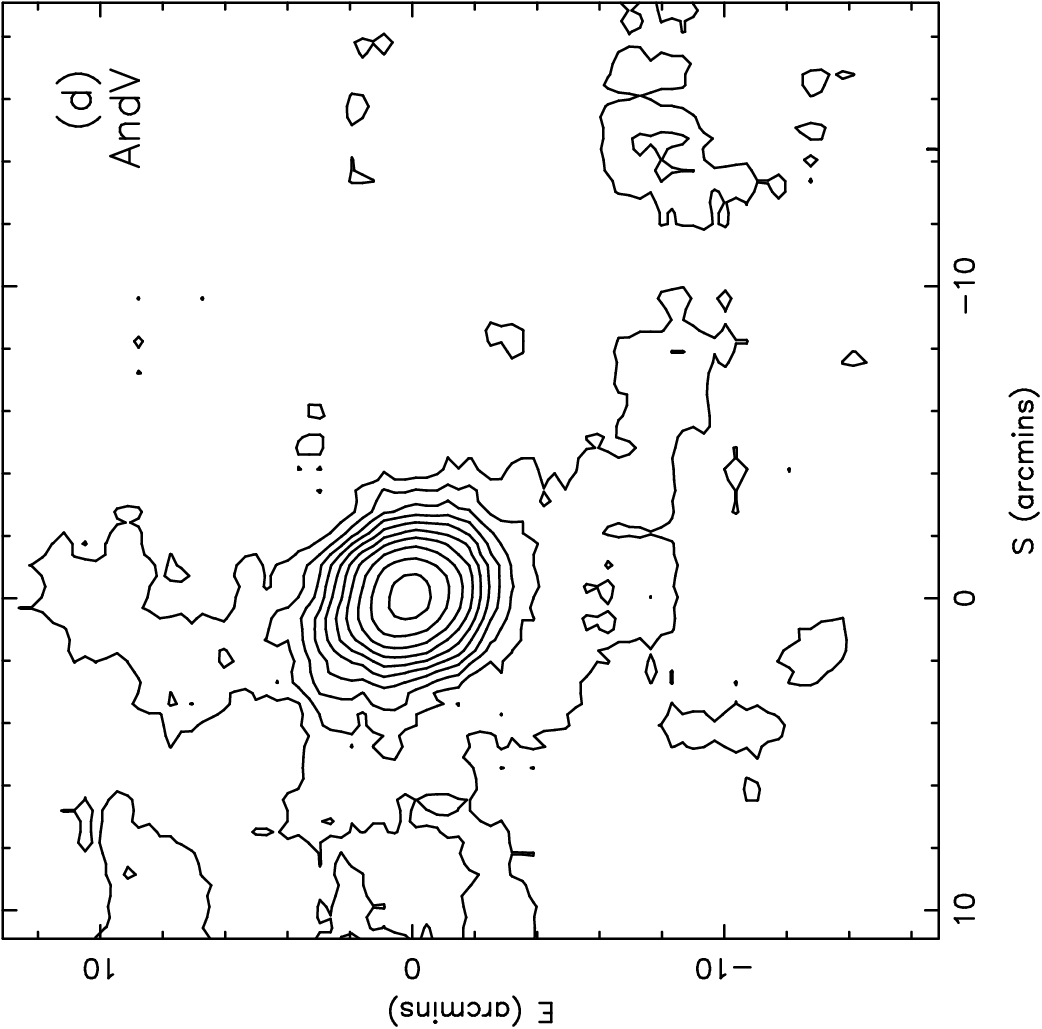}
\caption{Isopleth maps for the six dSph
satellites of M31 (panels (a) -- (f)), and the isolated dSph in Cetus
(g), in standard coordinate $\left(\xi, \eta\right)$ projections. Each
map shows a similar $0\degg5 \times 0\degg5$ area of sky. A large
range in structural properties is clearly evident for these
galaxies. The first contour of each map is $2\,\sigma$ above
background. Contour increments then increase by a factor $n$ for each
subsequent contour: (a) Andromeda~I: $n = 1.15$ (b) Andromeda~II: $n =
1.3$ (c) Andromeda~III: $n = 1.15$ (d) Andromeda~V: $n = 1.15$ (e)
Andromeda~VI $n = 1.25$ (f) Andromeda~VII: $n = 1.33$ (g) Cetus: $n =
1.25$.}
\label{maps}
\end{center}
\end{figure*}
\addtocounter{figure}{-1}
\begin{figure*}
\begin{center}
\includegraphics[angle=270, width=7.cm]{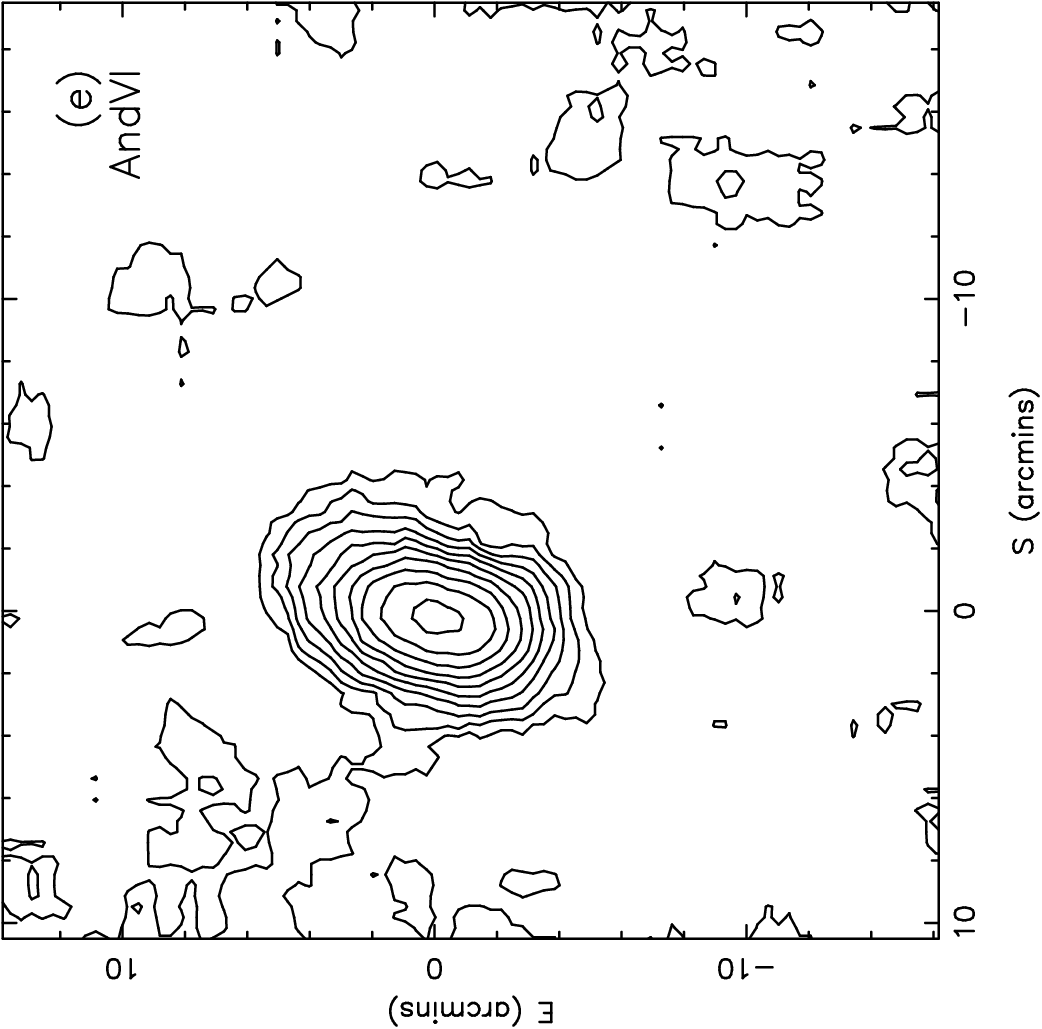}\hspace{5mm}
\includegraphics[angle=270, width=7.cm]{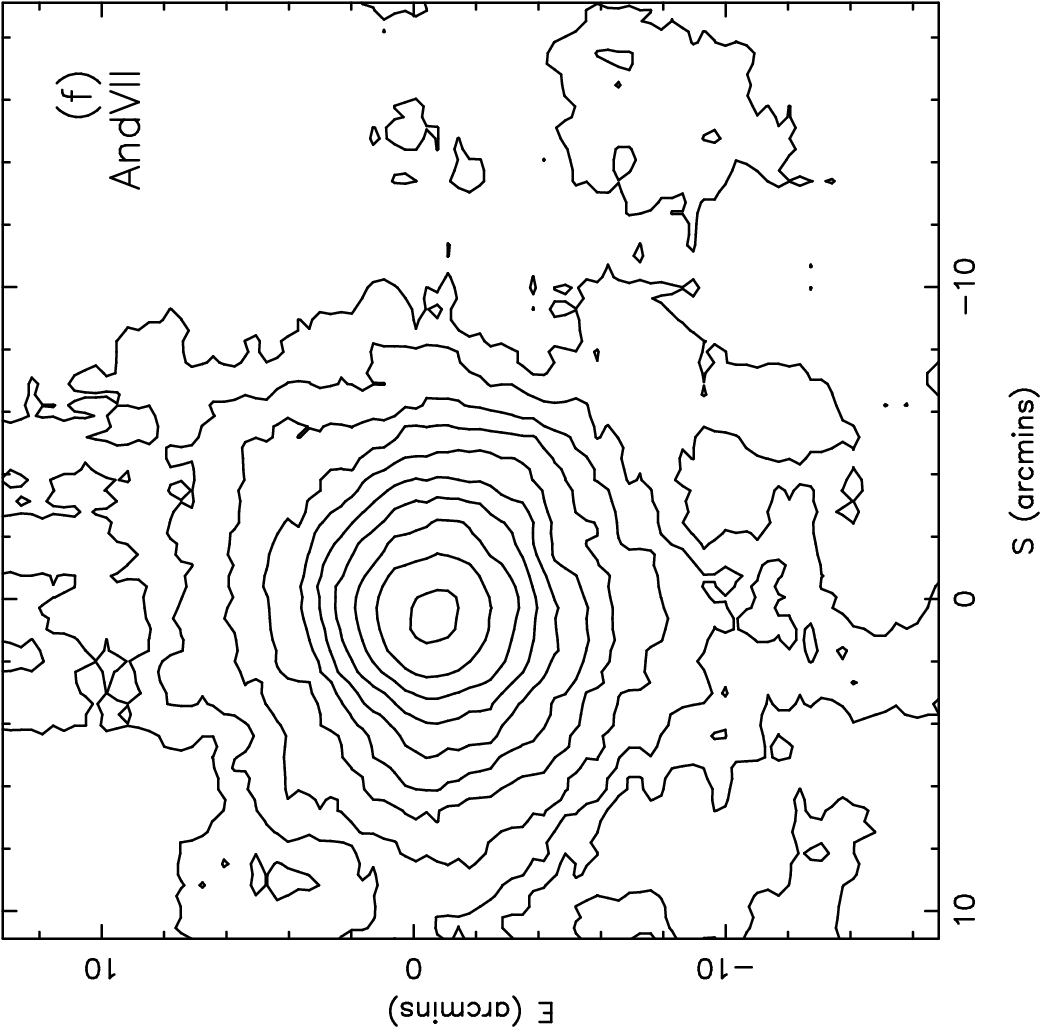}\vspace{5mm}
\includegraphics[angle=270, width=7.cm]{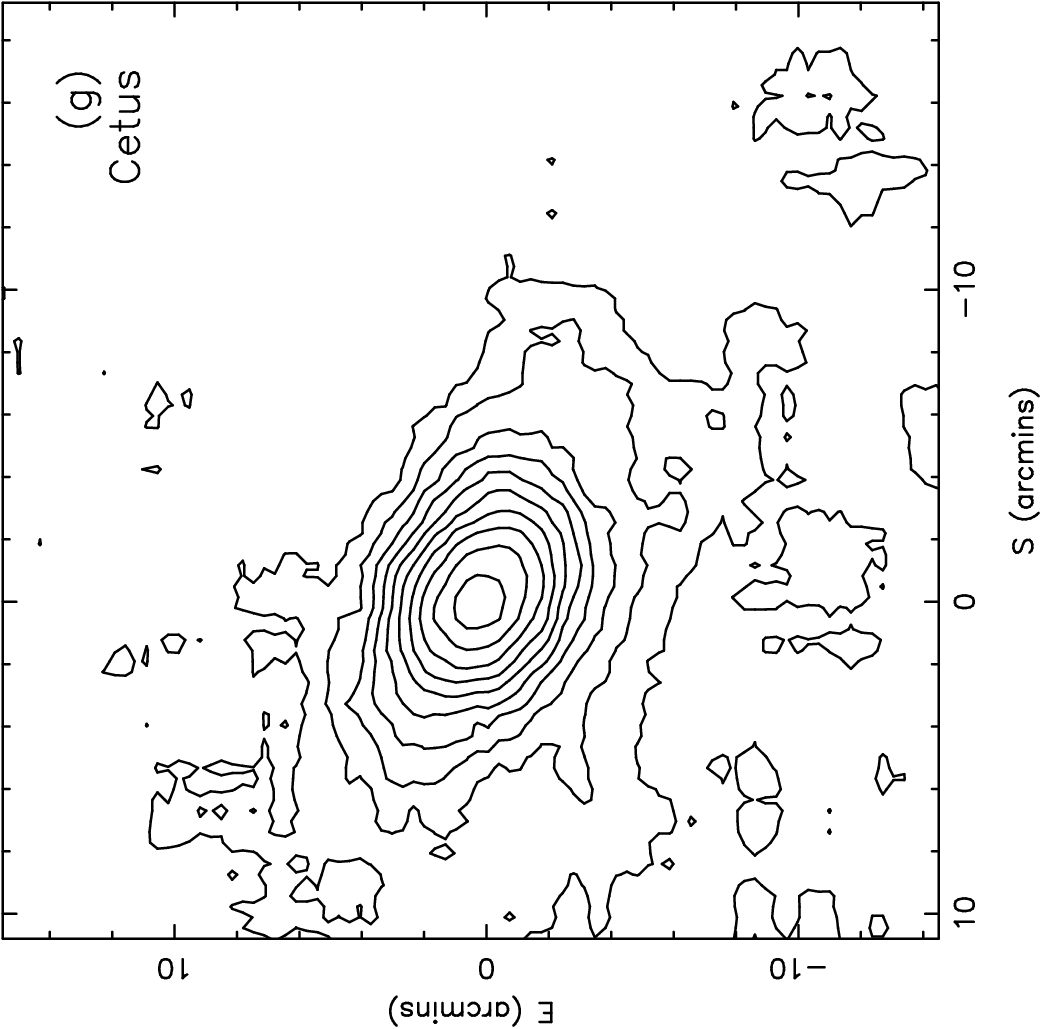}
\caption{{\it- continued.}}
\end{center}
\end{figure*}

\subsection{Radial profiles and associated parameters}
\label{radial}

The stellar radial profile for each galaxy is constructed by measuring
the stellar number density in elliptical annuli with the average
ellipticity and position angle derived above and applying the
background correction. Assuming fixed parameters for each galaxy gives
a robust estimate of the profile suitable for comparison with other
galaxies and models. The width of the elliptical annuli are set by
requiring a minimum signal-to-noise threshold in each bin. The results
are plotted in logarithmic form in Figure~\ref{sbprof}. The error bars
take into account the Poisson error in the counts and the uncertainty
in the background estimate.

\begin{figure*}
\begin{center}
\includegraphics[angle=0, width=7.5cm]{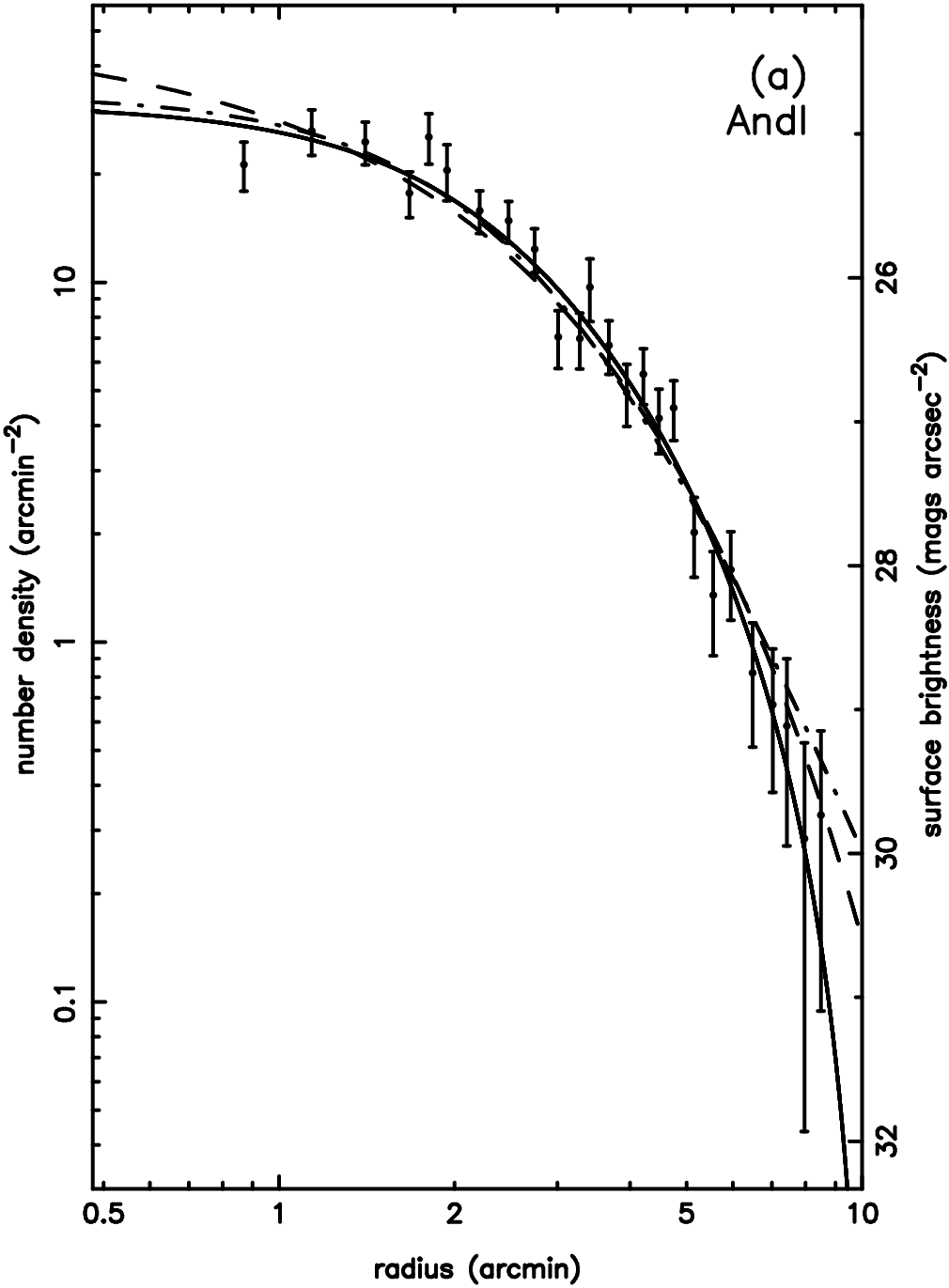}\hspace{5mm}
\includegraphics[angle=0, width=7.5cm]{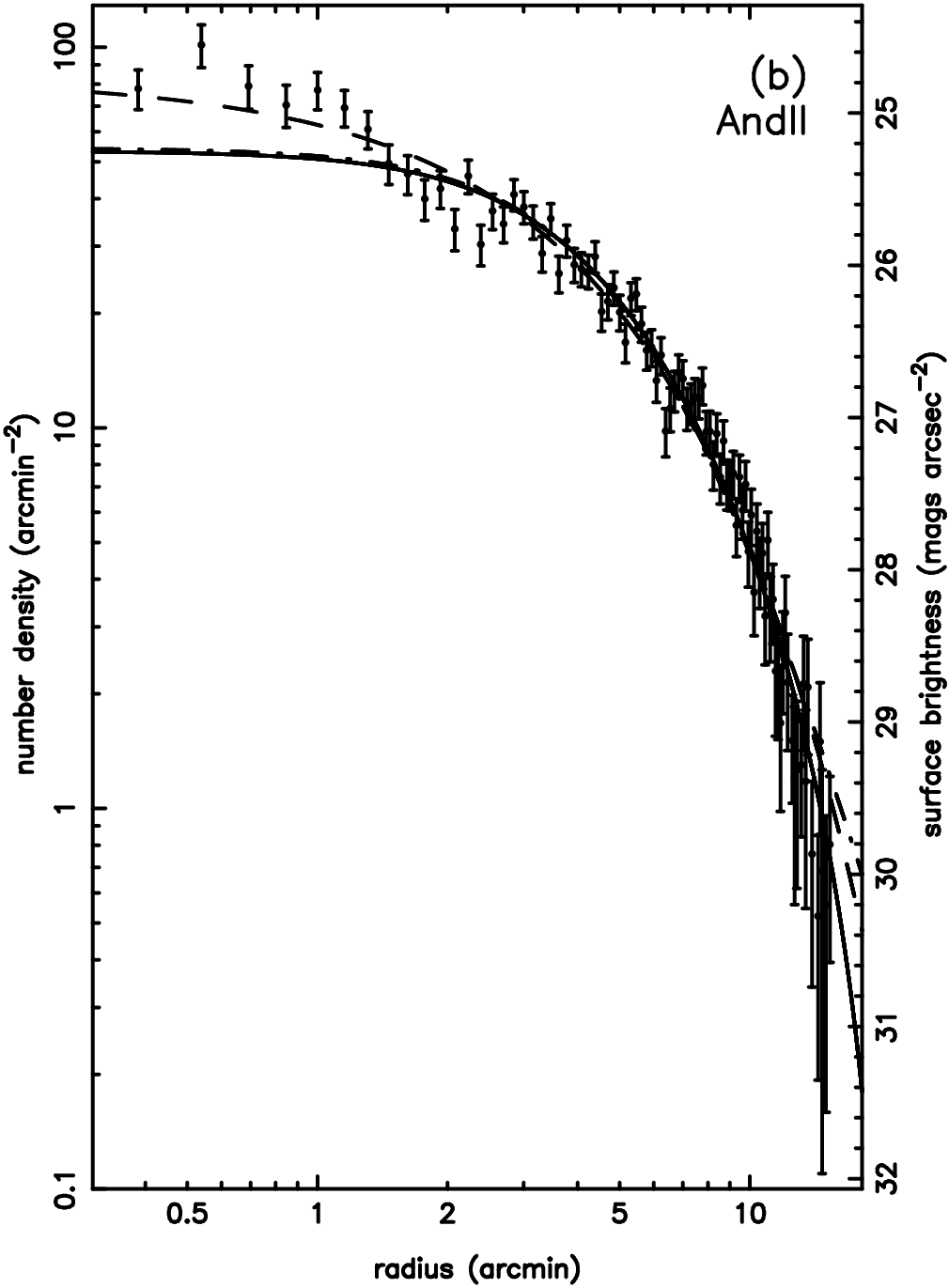}
\caption{Log-log plots of the stellar
radial profiles of the dSph galaxies. Error bars take into account the
Poisson uncertainty in the counts and the uncertainty in the
background estimate. The solid curves represent the best King profile
fit to the data (Equation~\ref{kingeqn}, Table~\ref{kingtab}), the
dashed lines correspond to the best exponential fit
(Equation~\ref{expeqn}, Table~\ref{exptab}) and the dot-dashed lines
represent the best Plummer model fit (Equation~\ref{plumeqn},
Table~\ref{exptab}). The surface brightness scale on the right
vertical axis has been calculated by normalising the data to the
integrated flux measurement in Section~\ref{integrate}.}
\label{sbprof}
\end{center}
\end{figure*}
\addtocounter{figure}{-1}
\begin{figure*}
\begin{center}
\includegraphics[angle=0, width=7.5cm]{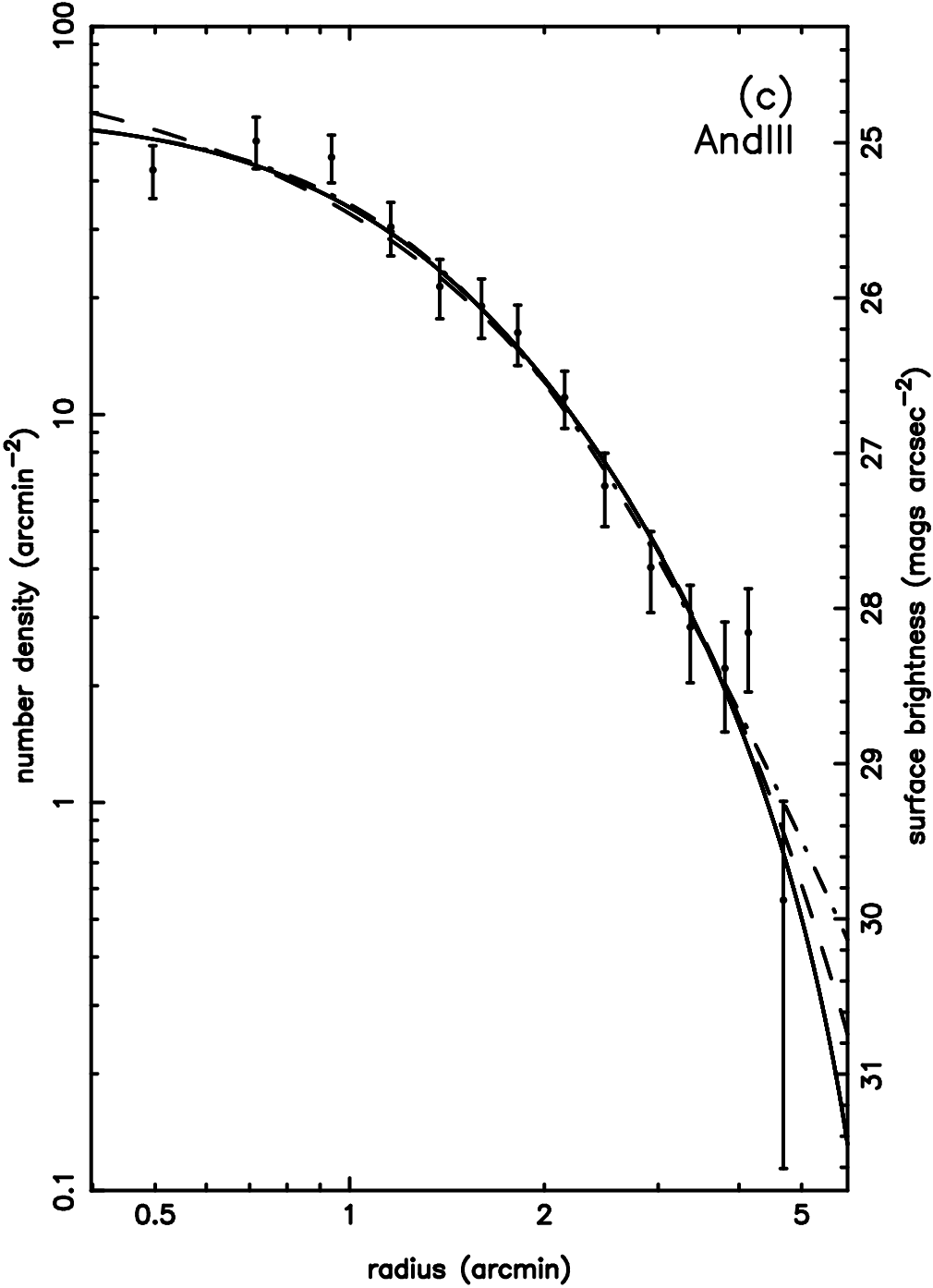}\hspace{5mm}
\includegraphics[angle=0, width=7.5cm]{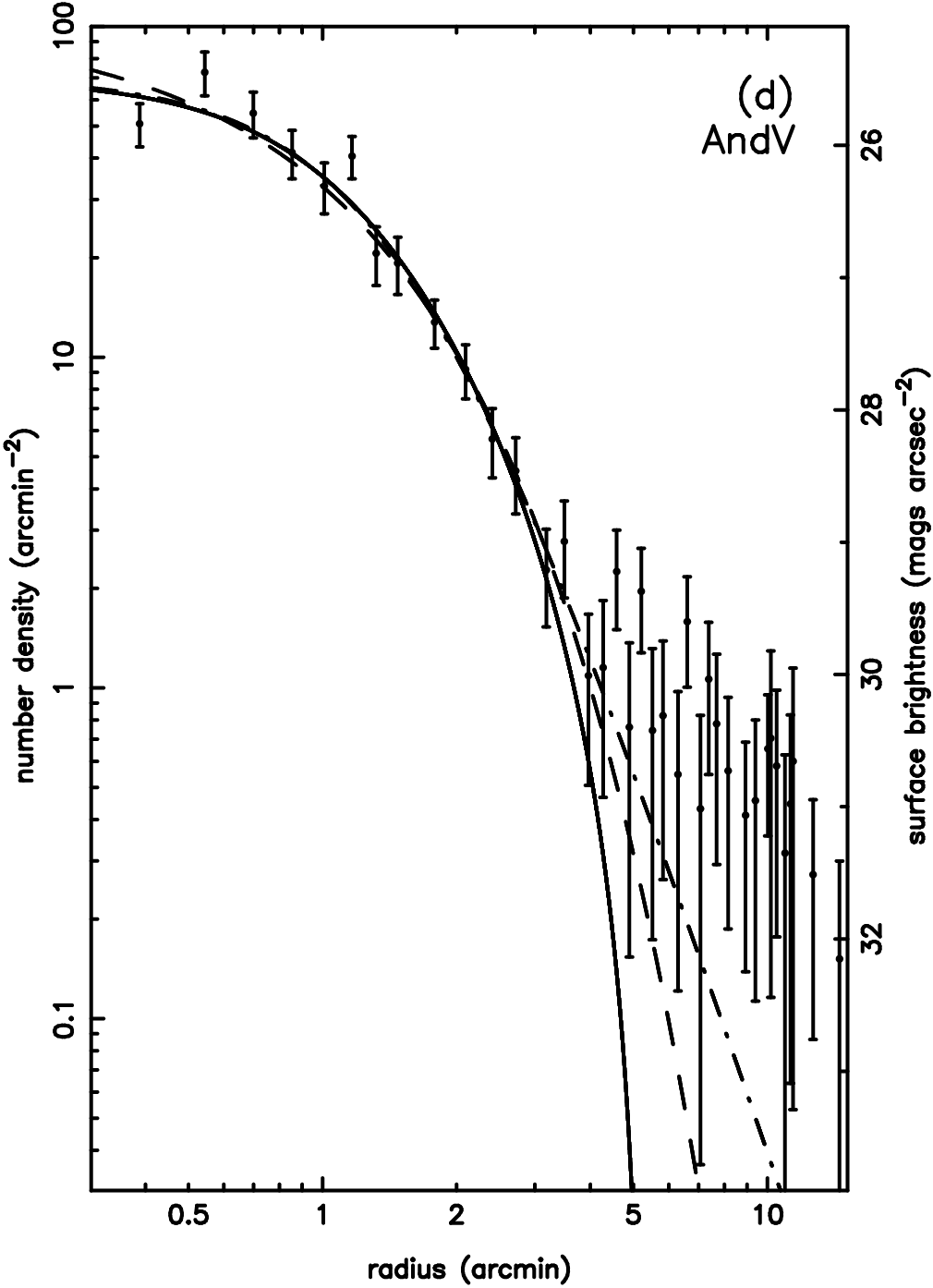}\vspace{5mm}
\includegraphics[angle=0, width=7.5cm]{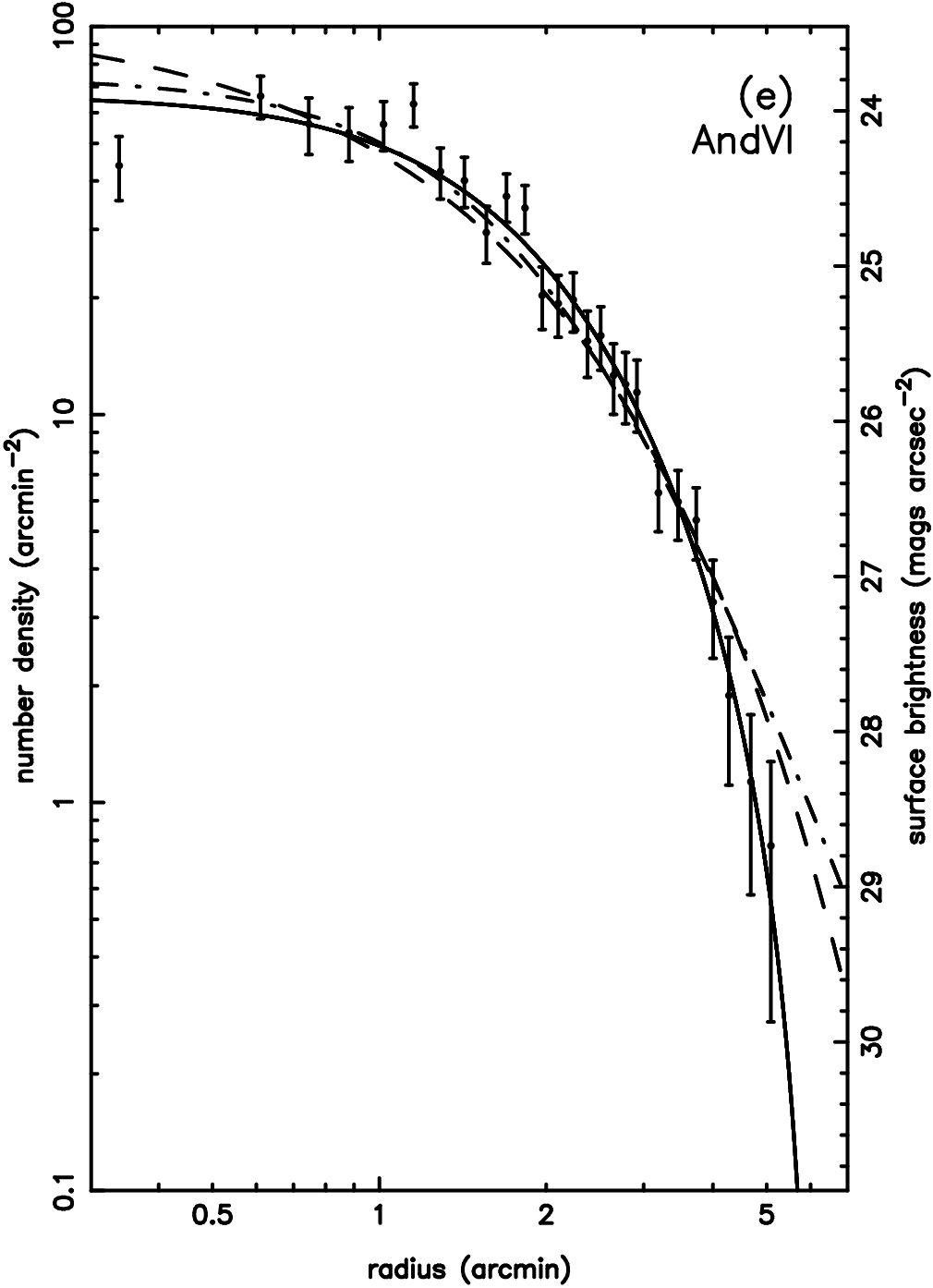}\hspace{5mm}
\includegraphics[angle=0, width=7.5cm]{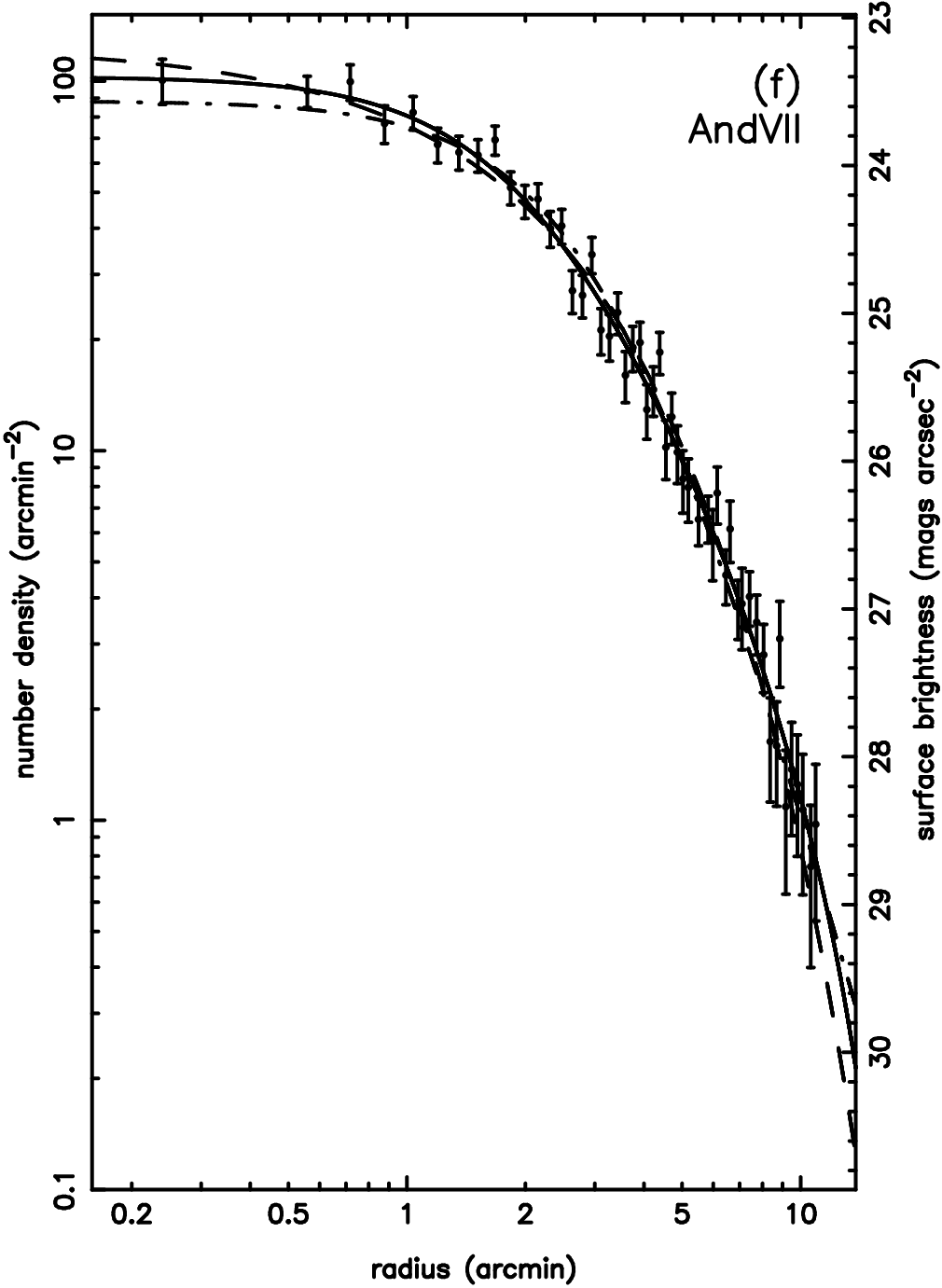}
\caption{{\it- continued.}}
\end{center}
\end{figure*}
\addtocounter{figure}{-1}
\begin{figure}
\begin{center}
\includegraphics[angle=0, width=7.5cm]{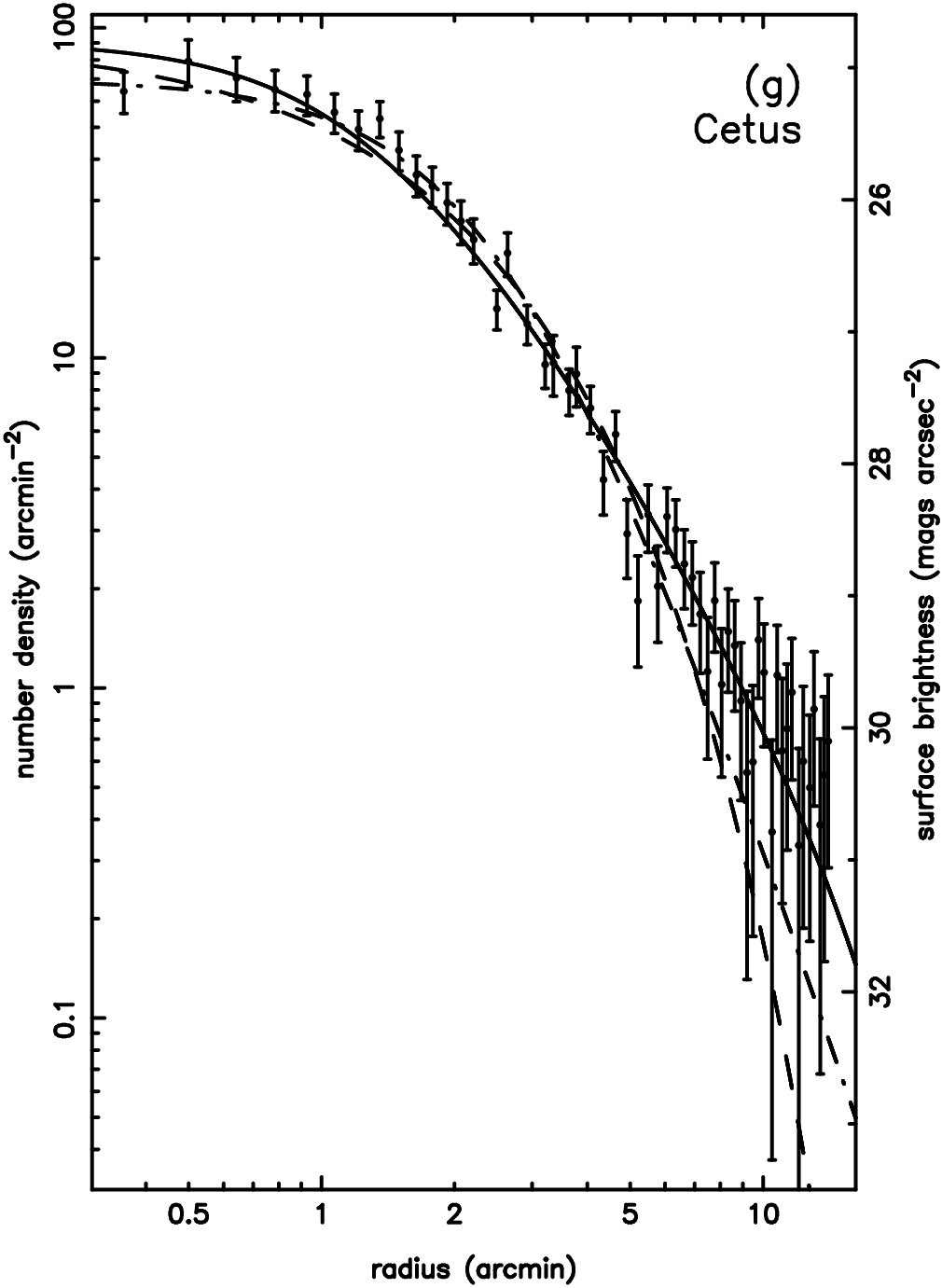}
\caption{{\it- continued.}}
\end{center}
\end{figure}

The expected form of the stellar density distribution in dSphs is
unknown, although several profiles have been suggested.  It is not our
intention to compare the radial distributions in Figure~\ref{sbprof}
to every model which has been proposed, and we  have instead adopted the
simplest, most intuitive and most commonly used. In particular, King
profiles are used and the simple, empirical form of these is given by

\begin{equation}
f_K = A \left(
\frac{1}{\left[1+\left(r/r_c\right)^2\right]^{\frac{1}{2}}} -
\frac{1}{\left[1+\left(r_t/r_c\right)^2\right]^{\frac{1}{2}}}\right)^2
\label{kingeqn}
\end{equation}

\noindent (\citealt{king1962}). $A$ is a scaling parameter, $r$ the
radius from the centre of the system, $r_c$ the core radius and $r_t$
the tidal radius.  For the range of core concentrations considered
here ($c = \log_{10}\left(r_t/r_c\right) < 1.5$), these profiles
closely match the physically-motivated form derived in \cite{king1966}
for globular clusters, which is based on a lowered Gaussian
distribution function (\citealt{binney1987}). King profiles - both the
empirical and physically-motivated forms - provide a tractable family
of models with intuitive parameters that have been fitted extensively
to dwarf galaxies (\citealt{hodge1966,eskridge1988b,eskridge1988a};
IH95). We therefore adopt them as one of the models to be compared to
the radial distributions in Figure~\ref{sbprof}, although we note that
there is not necessarily a physically-motivated reason to expect King
profiles to be a realistic form of the radial profile for dwarf
galaxies, and interpretations based on this assumption should be
treated with caution.

\cite{faber1983} advocate exponential profiles to describe the
projected surface density distribution of dSphs,

\begin{equation}
f_E = B \exp\left(-\frac{r}{r_e}\right)~,
\label{expeqn}
\end{equation}

\noindent where $B$ is a scaling parameter and $r_e$ is the effective
radius. These models require one parameter less than King models and
qualitatively often provide as good a fit as the more complex
family. \cite{read2004} have shown that exponential profiles are a
generic phase of dSph evolution if they undergo rapid mass loss at
early times. $r_e$ is a useful parameter and we additionally compare
this class of model to the stellar radial distributions.

Finally, Plummer models are frequently used in N-body simulations of
dwarf galaxy disruption in a tidal field
(e.g. \citealt{font2004,law2005}). They are of the form

\begin{equation}
f_P = C \frac{b^2}{\left(b^2 + r^2\right)^2}~,
\label{plumeqn}
\end{equation}

\noindent where $C$ is a scaling parameter and $b$ is the Plummer core
radius. Due to their common use in simulations, we adopt this as the
final model to compare with the dSph profiles.

These three models are fitted to each of the stellar radial profiles
using a least-squares minimisation technique. The best-fitting King,
exponential and Plummer models are overlaid on the logarithmic
profiles in Figure~\ref{sbprof} as the solid, dashed and dot-dashed
curves respectively. The derived parameters corresponding to each of
these models is listed for each galaxy in Tables~\ref{kingtab} and
\ref{exptab}, along with the associated uncertainties and the formal
value of the reduced $\chi^2$ statistic. The concentration parameter
for the King profile is also given. The scale-radii correspond to the
geometric mean along the two axes of the dwarf.  The surface
brightness scale on the right vertical axis of Figure~\ref{sbprof} has
been calculated by normalising the radial profile to the integrated
flux measurements described in the following section.

\begin{table*}
\begin{center}
\begin{tabular*}{0.75\textwidth}{r c c c c c c c}
\hline
    & \multicolumn{6}{c}{King Profile} \\
&$\chi^2$&$r_c$ (\,$^\prime$)&$r_t$ (\,$^\prime$)&$r_c$ (\,kpc)&$r_t$ (\,kpc) & $c = \log_{10}\left(r_t/r_c\right)$\\
\hline
&&&&&&\\
 Andromeda I    & 1.01  & 2.7 $\pm$ 0.3 & 10.4 $\pm$ 0.9 & 0.58 $\pm$ 0.06 & 2.3 $\pm$ 0.2 & 0.59 $\pm$ 0.06 \\
&&&&&&\\
 Andromeda II   & 1.84  & 5.2 $\pm$ 0.2 & 22.0 $\pm$ 1.0 & 0.99 $\pm$ 0.04 & 4.2 $\pm$ 0.2 & 0.63 $\pm$ 0.03 \\
&&&&&&\\
 Andromeda III  & 0.89  & 1.3 $\pm$ 0.2 &  7.2 $\pm$ 1.2 & 0.29 $\pm$ 0.04 & 1.5 $\pm$ 0.3 & 0.74 $\pm$ 0.10 \\
&&&&&&\\
 Andromeda V    & 1.02  & 1.2 $\pm$ 0.2 &  5.3 $\pm$ 1.0 & 0.28 $\pm$ 0.04 & 1.2 $\pm$ 0.2 & 0.63 $\pm$ 0.11 \\
&&&&&&\\
 Andromeda VI   & 0.96  & 2.1 $\pm$ 0.2 &  6.2 $\pm$ 0.4 & 0.48 $\pm$ 0.06 & 1.4 $\pm$ 0.1 & 0.46 $\pm$ 0.06 \\
&&&&&&\\
 Andromeda VII  & 0.91  & 2.0 $\pm$ 0.1 & 19.3 $\pm$ 1.6 & 0.45 $\pm$ 0.02 & 4.3 $\pm$ 0.4 & 0.98 $\pm$ 0.04 \\
&&&&&&\\
 Cetus          & 1.00  & 1.3 $\pm$ 0.1 & 32.0 $\pm$ 6.5 & 0.29 $\pm$ 0.02 & 7.1 $\pm$ 1.5 & 1.40 $\pm$ 0.10 \\
&&&&&&\\
\hline
\end{tabular*}
\caption{Details of the best-fitting King (Equation~\ref{kingeqn}) profiles shown in Figure~\ref{sbprof} as the solid curves. Each scale radius is the
  geometric mean for the dwarf. The distance moduli and associated
  uncertainties derived in McConnachie et al. (2005) have been used to
  transform units.}
\label{kingtab}
\end{center}
\end{table*}

\begin{table*}
\begin{center}
\begin{tabular*}{0.75\textwidth}{r c c c c c c c c }
\hline
    & &\multicolumn{3}{c}{Exponential Profile}& &\multicolumn{3}{c}{Plummer Profile}\\
&&$\chi^2$&$r_e$ (\,$^\prime$)&$r_e$ (\,kpc)&&$\chi^2$&$b$ (\,$^\prime$)&$b$ (\,kpc)\\
\hline
&&&&&&&&\\
 Andromeda I   && 1.36 & 1.72 $\pm$ 0.06 & 0.38 $\pm$ 0.02  && 1.22 & 3.12 $\pm$ 0.12 & 0.67 $\pm$ 0.03 \\
&&&&&&&&\\
 Andromeda II  && 1.66 & 3.53 $\pm$ 0.06 & 0.67 $\pm$ 0.01  && 2.14 & 6.44 $\pm$ 0.10 & 1.23 $\pm$ 0.02 \\
&&&&&&&&\\
 Andromeda III && 1.00 & 1.00 $\pm$ 0.05 & 0.24 $\pm$ 0.01  && 0.76 & 1.82 $\pm$ 0.10 & 0.40 $\pm$ 0.03 \\
&&&&&&&&\\
 Andromeda V   && 1.25 & 0.86 $\pm$ 0.05 & 0.20 $\pm$ 0.01  && 0.92 & 1.56 $\pm$ 0.08 & 0.35 $\pm$ 0.02 \\
&&&&&&&&\\
 Andromeda VI  && 2.20 & 1.20 $\pm$ 0.04 & 0.27 $\pm$ 0.01  && 1.56 & 2.15 $\pm$ 0.08 & 0.49 $\pm$ 0.02 \\
&&&&&&&&\\
 Andromeda VII && 1.14 & 2.00 $\pm$ 0.04 & 0.44 $\pm$ 0.01  && 1.18 & 3.47 $\pm$ 0.07 & 0.77 $\pm$ 0.02 \\
&&&&&&&&\\
 Cetus         && 2.06 & 1.59 $\pm$ 0.05 & 0.34 $\pm$ 0.02  && 1.56 & 2.69 $\pm$ 0.08 & 0.59 $\pm$ 0.02 \\
&&&&&&&&\\
\hline
\end{tabular*}
\caption{Details of the best-fitting exponential
(Equation~\ref{expeqn}, dashed curves in Figure~\ref{sbprof}) and
Plummer (Equation~\ref{plumeqn}, dot-dashed curves in
Figure~\ref{sbprof}) models. Each scale radius is the geometric mean
for the dwarf. The distance moduli and associated uncertainties
derived in McConnachie et al. (2005) have been used to transform units.}
\label{exptab}
\end{center}
\end{table*}

\section{Integrated Photometry}
\label{integrate}

We have directly estimated the central surface brightness and integrated 
luminosity of the dSphs from the integrated flux distribution of each
galaxy.  With suitable image processing, the effects of Galactic
foreground stars and random noise can be reduced to manageable levels,
enabling these integrated parameters to be measured in a simple manner.

The processing procedure is relatively straightforward. First, the
existing object catalogues are used to define a bright foreground star
component at least 0.5 -- 1\,mags above the TRGB, to allow for the
potential presence of AGB stars. A circular aperture is then excised
around each foreground star and the flux within this aperture is set
to the local sky level, interpolated from a whole-frame background
map. The size of this aperture is the maximum of four times the
catalogue-recorded area of the bright star at the detection isophote,
or a diameter four times the derived FWHM seeing. Each frame is then
re-binned on a $3 \times 3$ grid to effectively create 1 arcsec
pixels. The binned image is then further smoothed using a 2D Gaussian
filter with a FWHM of 5\,arcsecs.

The result of this procedure is to produce a coarsely sampled smooth
image containing both the resolved and unresolved light contribution
from the dSph.  The central surface brightness can then be trivially
measured by deriving the radial profile, here defined as the median
flux value within elliptical annuli.  Finally, large elliptical
apertures are placed over the dSph and several comparison regions to
estimate the background-corrected integrated flux from the dwarf and
the reference regions.  The variation in the flux from the multiple
comparison measures gives a good indication of the flux error, which
is, of course, dominated by systematic fluctuations rather than by
random noise.  To mitigate the effect of random residual foreground
stellar halos and scattered light from bright stars just outside the
field of view, the elliptical apertures are chosen to correspond to
the derived value of $r_{\frac{1}{2}}$ for the dSph.  The estimated
total flux is then scaled to allow for this correction. For
Andromeda~VII, a quarter-light radius is used instead as there is a
significant Galactic nebulosity in this field which makes estimates
based on larger apertures unreliable.  As with the resolved component,
some of the comparison regions may contain light from the
dSph. However, this effect is negligible in comparison to the
systematic fluctuation.

\begin{table*}
\begin{center}
\begin{tabular*}{0.65\textwidth}{r c c c c c c c c}
\hline
    & \multicolumn{8}{c}{Integrated Photometry} \\
&& $m_V$ & $M_V$ & $V_o$ & $S_o$ & $r_{\frac{1}{2}}~\left(^\prime\right)$& $r_{\frac{1}{2}}~\left({\rm kpc}\right)$& \\
\hline
&&&&&&&&\\
 Andromeda I   && 12.7 $\pm$ 0.1 & -11.8 $\pm$ 0.1 & 24.7 & 24.8 & 2.8 & 0.60 &\\
&&&&&&&&\\
 Andromeda II  && 11.7 $\pm$ 0.2 & -12.6 $\pm$ 0.2 & 24.5 & 24.7 & 5.6 & 1.06 &\\ 
&&&&&&&&\\
 Andromeda III && 14.4 $\pm$ 0.3 & -10.2 $\pm$ 0.3 & 24.8 & 24.7 & 1.6 & 0.36 &\\
&&&&&&&&\\
 Andromeda V   && 15.3 $\pm$ 0.2 &  -9.6 $\pm$ 0.3 & 25.3 & 25.6 & 1.3 & 0.30 &\\  
&&&&&&&&\\
 Andromeda VI  && 13.2 $\pm$ 0.2 & -11.5 $\pm$ 0.2 & 24.1 & 23.9 & 1.8 & 0.42 &\\ 
&&&&&&&&\\
 Andromeda VII && 11.8 $\pm$ 0.3 & -13.3 $\pm$ 0.3 & 23.2 & 23.6  & 3.3 & 0.74 &\\
&&&&&&&&\\
 Cetus         && 13.2 $\pm$ 0.2 & -11.3 $\pm$ 0.3 & 25.0 & 24.8 & 2.7 & 0.60 &\\        
&&&&&&&&\\
\hline
\end{tabular*}
\caption{The integrated photometric properties of
the dSphs. The integrated apparent magnitude ($m_V$) is uncorrected
for reddening, while the absolute magnitudes ($M_V$) and central
surface brightnesses have been de-reddened. $V_o$ is the directly measured
central surface brightness while $S_o$ is the central surface
brightness obtained by normalising the radial profile to the
integrated luminosity of the dSph. The values for $r_{\frac{1}{2}}$ are obtained
by integration of the appropriate King profile, with the exception of
Andromeda~II where a two-component model has been used (see
Section~\ref{m31dsphs}), and Cetus, where the exponential profile has
been used.}
\label{inttab}
\end{center}
\end{table*}

The results for each galaxy are listed in Table~\ref{inttab}. The
integrated apparent magnitude has not been corrected for extinction,
but all other quantities are extinction-corrected using the values
given by \cite{schlegel1998}, tabulated in Table~1 of
\cite{mcconnachie2005a}. The distance measurements and associated
uncertainties derived in \cite{mcconnachie2005a} have also been used.
$V_o$ is the directly measured central surface brightness for the dSphs. $S_o$
is the derived value of the central surface brightness obtained by
normalising the King profile such that the integral under the profile
is equal to the integrated magnitude for the dSph. For Andromeda~II, a
two-component profile has been used instead of the King profile (see
Section~\ref{m31dsphs}). The uncertainty on the measurement of $S_o$
is dominated by the uncertainty on the absolute magnitude
for the dSph, and the uncertainty on the directly measured central 
surface brightness $V_o$ is estimated to be $\sim
0.1 - 0.2$\,mags. There is reasonable agreement between the values for
$V_o$ and $S_o$, which highlights the accuracy and consistency of the
integrated magnitude measurement, the radial profiles, and the surface
brightnesses.

Table~\ref{inttab} also lists the derived values of $r_{\frac{1}{2}}$
of each dSph, obtained by integration of the appropriate King profile.
This is with the exception of Andromeda~II, where a two-component
profile was used, and Cetus, where $r_{\frac{1}{2}}$ has been derived
by integration of the exponential profile. The lack of an obvious
truncation radius in the Cetus data results in a value for
$r_{\frac{1}{2}}$ derived from the King profile, which is twice as
large as from either the Plummer or exponential profiles, and which is
probably unreliable. The half-light radius of a Plummer profile is
equal to $b$, and the half-light radius of an exponential profile is
$1.68\,r_e$. Generally, these are consistent with the tabulated value
for $r_{\frac{1}{2}}$ measured from the King model fit.

\section{Discussion}
\label{compare}

Figures~\ref{maps} and \ref{sbprof}, and Tables~\ref{geom},
\ref{kingtab}, \ref{exptab} and \ref{inttab} present many results
relating to the Cetus and M31 dSphs. In this section, we compare these
results to those of previous studies (Section~\ref{previous}) and then
discuss each dSph individually (Sections~\ref{cetusdsph} and
\ref{m31dsphs}). We then compare the M31 and Galactic populations
(Section~\ref{differences}) and examine correlations in the Local Group
population as a whole (Section~\ref{similarities}).

\subsection{Comparison with previous work}
\label{previous}

\cite{caldwell1992} and \cite{caldwell1999} derived structural
parameters for Andromeda I, II, III, V, VI and VII based upon the
integrated light of these galaxies. \cite{whiting1999} has derived the
structural parameters for Cetus. Table~\ref{prevtab} lists the values
for $r_{\frac{1}{2}}$, $m_V$ and $M_V$ (corrected for the extinction measurements
and distances used here) from \cite{caldwell1992} (Andromeda I, II and
III), \cite{caldwell1999} (Andromeda V, VI and VII), and
\cite{whiting1999} (Cetus). The value of $r_{\frac{1}{2}}$ for Cetus has been
evaluated by integrating the King profile in
\cite{whiting1999}. 

\begin{table*}
\begin{center}
\begin{tabular*}{0.5\textwidth}{rcccl@{\extracolsep{0mm}}c@{\extracolsep{0mm}}lcr }
\hline
    && $r_{\frac{1}{2}}~\left(^\prime\right)$ &&   & $m_V$ & && $M_V$     \\
\hline
&&&&&&\\
 Andromeda I   && 2.5 && 12.75 & $\pm$ & 0.07 && $-11.8$ \\
&&&&&&\\
 Andromeda II  && 2.3 && 12.71 & $\pm$ & 0.16 && $-11.6$ \\
&&&&&&\\
 Andromeda III && 1.3 && 14.21 & $\pm$ & 0.08 && $-10.4$ \\
&&&&&&\\
 Andromeda V   && 0.6 && 15.92 & $\pm$ & 0.14 &&  $-8.9$ \\
&&&&&&\\
 Andromeda VI  && 1.4 && 13.30 & $\pm$ & 0.12 && $-11.4$ \\
&&&&&&\\
 Andromeda VII && 1.3 && 12.90 & $\pm$ & 0.27 && $-12.2$ \\
&&&&&&\\
 Cetus         && 1.4 && 14.4  & $\pm$ & 0.2  && $-10.1$ \\
&&&&&&\\
\hline
\end{tabular*}
\caption{Previous estimates of $r_{\frac{1}{2}}$, $m_V$
and $M_V$ for the Cetus and M31 dSphs, taken from Caldwell et
al. (1992) (Andromeda I, II and III), Caldwell (1999) (Andromeda V, VI
and VII), and Whiting et al. (1999) (Cetus). The values for $M_V$ have
been recalculated for the distance and reddening estimates used here.}
\label{prevtab}
\end{center}
\end{table*}

We measure $r_{\frac{1}{2}}$ for Cetus to be twice that of
\cite{whiting1999}. These authors trace the Cetus surface brightness
profile to $\sim 4$\,arcmins, while the measurement here extends to
$\sim 12$\,arcmins. Additionally, the limiting surface brightness of
\cite{whiting1999} is $\sim 29.5$\,mags\,arcsec$^{-2}$ (see their
Figure~4), while our profile extends $\sim 1.5$\,mags\,arcsec$^{-2}$
deeper. As a result, we find Cetus extends much further than originally
measured. The brighter luminosity we derive is fully consistent with
this increase in size. Identical arguments apply for Andromeda~V and
VII, where we measure their radial profiles out to much larger radii
and over a larger range in surface brightness than
\cite{caldwell1999}, and find them to be more extended and therefore brighter.

A similar reason to the one above is not responsible for the
difference in results for Andromeda~II. \cite{caldwell1992} measures
the surface brightness profile for this galaxy over a similar radial
and surface brightness range as we do, yet we derive a value for
$r_{\frac{1}{2}}$ that is more than twice as large as that derived by
\cite{caldwell1992}. We believe this is a result of the complexity of
the profile of Andromeda~II, which we discuss in Section~4.3.
 
\subsection{The isolated dwarf spheroidal in Cetus}
\label{cetusdsph}

In the Local Group, there is a preference for dSph galaxies to be
found as satellites of either M31 or the Galaxy, whereas dwarf
irregular galaxies are preferentially found in more isolated
environments. The origin of this position -- morphology relation is
not yet known, although it may well be related to the ram pressure and
tidal stripping of dwarf galaxies near large galaxies by hot, gaseous
haloes (\cite{mayer2001a,mayer2001b,mayer2005}). Cetus and Tucana are
the only dSph galaxies in the Local Group which are not clearly
satellites of either M31 or the Galaxy. Analysis of their properties,
and comparison to the dSphs which are satellites, may prove useful in
determining the origin of this position -- morphology relation and in
determining which evolutionary processes can be attributed to
environmental effects.

The King model tidal radius of Cetus is $r_t = 6.6$\,kpc, making it 
apparently the largest dSph in the Local Group by a significant amount.
However, the stellar radial profile out to 12\,arcmins does not show
any evidence of turning over, and the derived value of $r_t$ is well
beyond the final data point. Therefore, the concept of a tidal radius
in this case is probably misleading as Cetus shows no evidence of
tidal truncation. Cetus is currently $\simeq 750$\,kpc from the
Galaxy, which places it $\simeq 680$\,kpc from M31. If Cetus has spent
most of its evolution in isolation then we would not expect it to be
truncated; its profile is fully consistent with an undisturbed system.
From Equation~7. and assuming the masses of M31 and the Galaxy to be
$\simeq 10^{12}\,M_\odot$, then if Cetus has a typical mass for a dSph
($\sim 5 \times 10^7\,M_\odot$), the lack of a tidal radius out
to 6 or 7\,kpc implies that after its last major star formation episode,
Cetus has never been much closer that
$\sim 200$\,kpc to either M31 or the Galaxy. 

Kinematic information for Cetus is required to compliment this data. A
radial velocity measurement may show whether the motion of Cetus
suggests it ever having interacted with the Galaxy or M31. In
addition, a stellar velocity dispersion measurement will constrain its
mass. This information will help decide whether Cetus poses a
challenge for formation and gas-loss models of dSphs which currently
depend upon interactions with larger systems.

\subsection{The M31 dwarf spheroidals}
\label{m31dsphs}

{\bf Andromeda~I} is one of the largest and brightest M31 dSphs and
displays the distinctive S-shape indicative of ongoing tidal
disruption (Figure~\ref{maps}a), suggesting the presence of low
surface brightness tidal tails. Although the average position angle of
Andromeda~I is $\sim 22^\circ$, its isophotes twist from $\sim
0^\circ$ in the centre to $\sim 40^\circ$ in the outer regions, again
indicating strong tidal disruption (\citealt{choi2002}). From the
surface brightness profile alone (Figure~\ref{sbprof}a) there is no
sign that Andromeda~I is being disrupted; there is no indication of a
break in the profile, whose presence would normally be interpreted as
due to tidal heating/disruption (\citealt{choi2002}). Of course, this
signature might be found at a lower surface brightness threshold.

Andromeda~I shows the strongest evidence of tidal disruption of all
the M31 dSphs from its contours alone. Deeper photometry and
kinematics of this object will reveal the full extent of its tidal
disruption. Kinematic data will be particularly interesting, as it is
possible that Andromeda~I is surrounded by only a relatively shallow
dark matter halo. Otherwise, it is not obvious that the outer stellar
regions should be so distorted by the M31 tidal field.

{\bf Andromeda~II}, shown in Figure~\ref{maps}b, is huge and
circular. It is currently $\simeq 185$\,kpc from M31 and has a
 large King tidal radius ($r_t = 4.2$\,kpc). If its orbit is relatively
circular, then this could account for its appearance since the M31
tidal field at this distance should be small. The average ellipticity
of Andromeda~II is $\simeq 0.22$, although its central regions are
significantly more circular than this; the innermost three isophotes of
Figure~\ref{sbprof}b are visibly more ``core-like'' than its outer
regions.

The radial profile of Andromeda~II (Figure~\ref{sbprof}b) is quite
unusual. The formal reduced $\chi^2$ value for the King, Plummer and
exponential models are relatively poor; both the King and exponential
profiles underestimate the surface brightness in the central few
arcmins by nearly a factor of two, while the Plummer profile also
underestimates the central surface brightness slightly while failing
to match at intermediate radii. The radius at which the King profile
diverges from the data is close to the radius where the ``core''
develops in the isopleth map.

\begin{figure}
\begin{center}
\includegraphics[angle=0, width=7.5cm]{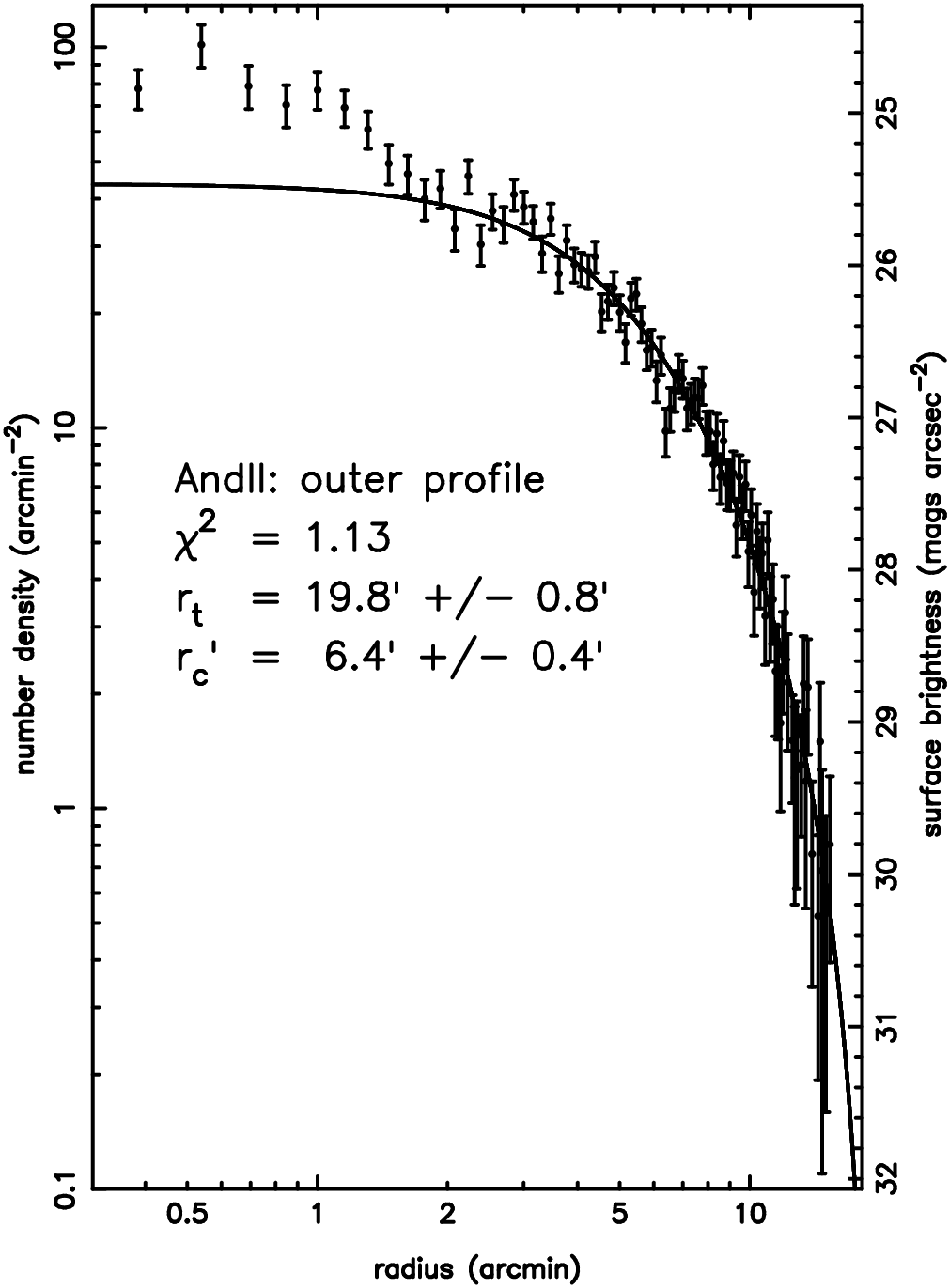}
\caption{The best fit King model to the
surface brightness profile of Andromeda~II, ignoring the inner
2\,arcmins. The formal value of the fit is good, and the divergence of
the King profile to the data at small radii suggests the presence of a
secondary core component.}
\label{andiiouter}
\end{center}
\end{figure}

Figure~\ref{andiiouter} is a King profile fit to the Andromeda~II
profile ignoring the inner $2$\,arcmins. The formal quality of the fit
is much better in this range ($\chi^2 = 1.13$) and diverges at the
same radius at which the ``core'' is observed in Figure~\ref{maps}b.
This reveals a factor of two excess of stars at small radius, which is
compelling evidence for a second component, in the form of a roughly
constant density stellar core. Deeper global photometry down to the
level of the horizontal branch and more kinematic information is
required to confirm this hypothesis and place the structure of this
galaxy in an evolutionary context. For example, it could be that this
represents a dissolved star cluster, or the secondary component may
reflect spatial variations in the star formation history, such that
there were two main episodes where one led to a more centrally
concentrated stellar population than the other. Indeed, from their HST
study of this galaxy, \cite{dacosta1996} are unable to model its
abundance distribution without assuming a two component
(``metal-rich'' and ``metal-poor'') chemical enrichment model. The
above scenarios, however, have only recently started to be considered
for the Galactic dSphs, motivated by the discoveries by
\citealt{kleyna2004} and \cite{tolstoy2004}.

{\bf Andromeda~III} is the most elongated dSph in this sample. Its
outer isophote in Figure~\ref{maps}c suggests that the outer regions
of this galaxy may be somewhat stretched and perturbed, although there
is no evidence for this in the surface brightness profile
(Figure~\ref{sbprof}c). Deeper photometry should reveal the true
extent of this sparse population. Andromeda~III is one the closest
dSphs to M31, at a distance of $\simeq 75$\,kpc.

{\bf Andromeda~V} ($M_V \simeq -9.6$\,mags) is a small and faint
dSph. The main body of the dwarf is fairly round and compact, but its
outermost isophotes appear diffuse and fuzzy (Figure~\ref{maps}d). It
is possible that this is due to heating of the outer region by the
tidal field of M31. The radial profile of Andromeda~V appears to show
a break at $3.5$\,arcmins, resulting in a failure of all three
parametric fits to follow the profile. Simulations show that tidal
effects can manifest themselves in the form of a break in the surface
brightness profile (\citealt{choi2002}), and so Andromeda~V may be
showing evidence of tidal harassment.

It is also worth noting that the absolute magnitude of Andromeda~V
which we derive is broadly consistent with a metallicity of [Fe/H]
$\simeq -1.5$ (\citealt{armandroff1998,mcconnachie2005a}).
Andromeda~V was previously measured to be fainter (Section~4.1), and
its original metallicity measurement of [Fe/H] $\simeq -1.5$ by
\cite{armandroff1998} meant that this object was an outlier on the
luminosity -- metallicity relation first shown by \cite{caldwell1999}.
A later measurement marked this object as more metal poor, at [Fe/H]
$\simeq -2.2$ (\citealt{davidge2002}), more consistent with its
low-luminosity measurement.  However, the most recent measurement of
its metallicity by \cite{mcconnachie2005a} agrees with the original
measurement of \cite{armandroff1998}. Given its absolute magnitude
measured here and its higher metallicity measurement, Andromeda~V
appears to follow the luminosity -- metallicity relation of
\cite{caldwell1999}, while its earlier discrepancy can be attributed
to an underestimated luminosity.

{\bf Andromeda~VI} has a radial profile that is well described by a
King profile. It shows no evidence of tidal disruption in either the
isopleth map (Figure~\ref{maps}e) or the surface brightness profile
(Figure~\ref{sbprof}e). Its $r_{\frac{1}{2}}$ is average for the M31
dSphs, and its $M_V$ and central surface brightness are likewise
relatively typical of the population.

{\bf Andromeda~VII} is the brightest and most extended of the M31
dSphs. In a similar way to Andromeda~II, Figure~\ref{maps}e shows it
to be nearly circular in projection, with an ellipticity of 0.13 and
$r_t = 4.3$\,kpc. Andromeda~VII is currently $\sim 220$\,kpc from M31
and a large tidal radius is not unexpected if its orbit is relatively
circular. Unlike Andromeda~II, its surface brightness profile shows no
deviation from a King model fit.

\subsection{Contrasting the M31 and Galactic dwarf spheroidals}
\label{differences}

\begin{figure*}
\begin{center}
\includegraphics[angle=270, width=7.5cm]{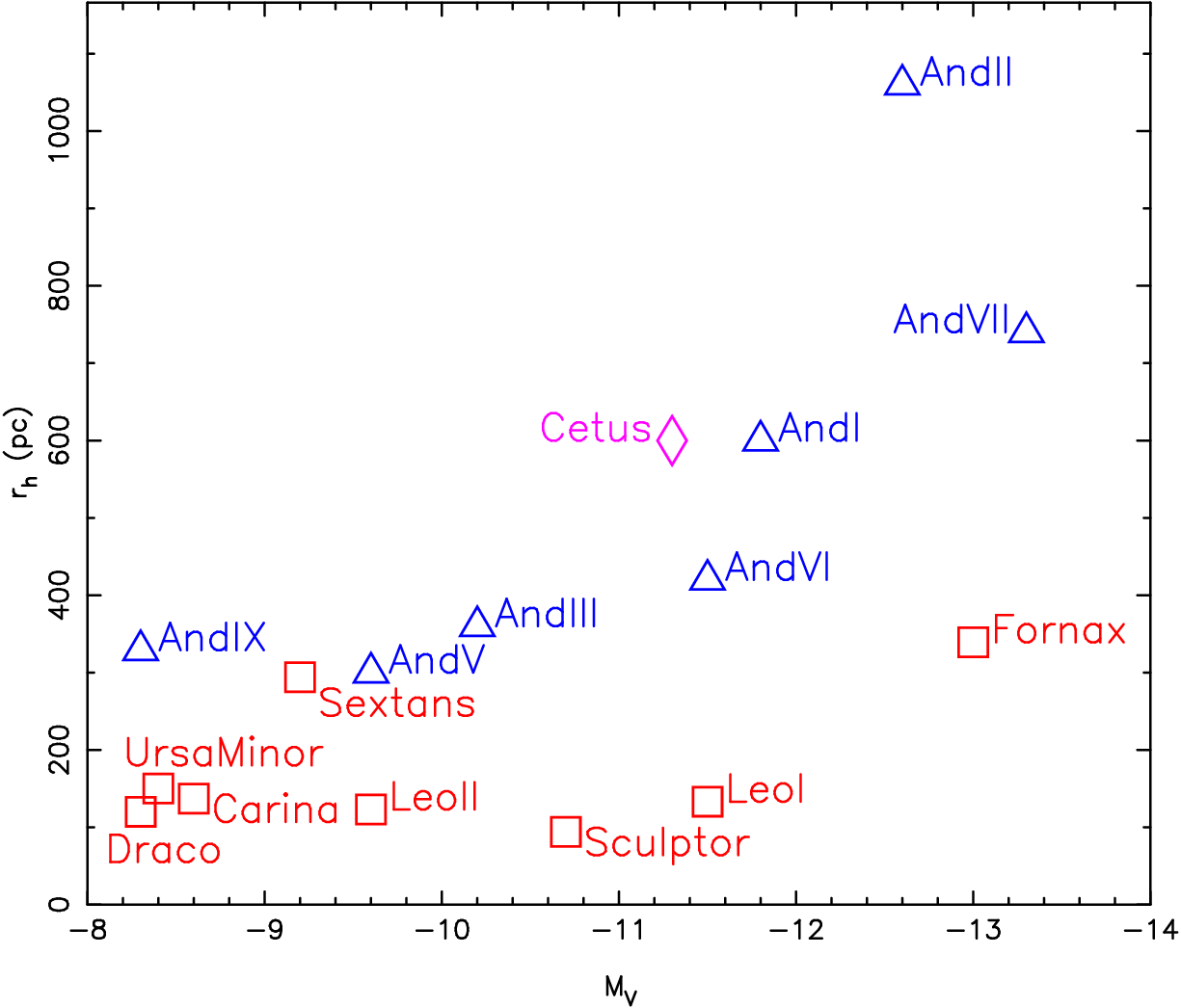}\hspace{5mm}
\includegraphics[angle=270, width=7.5cm]{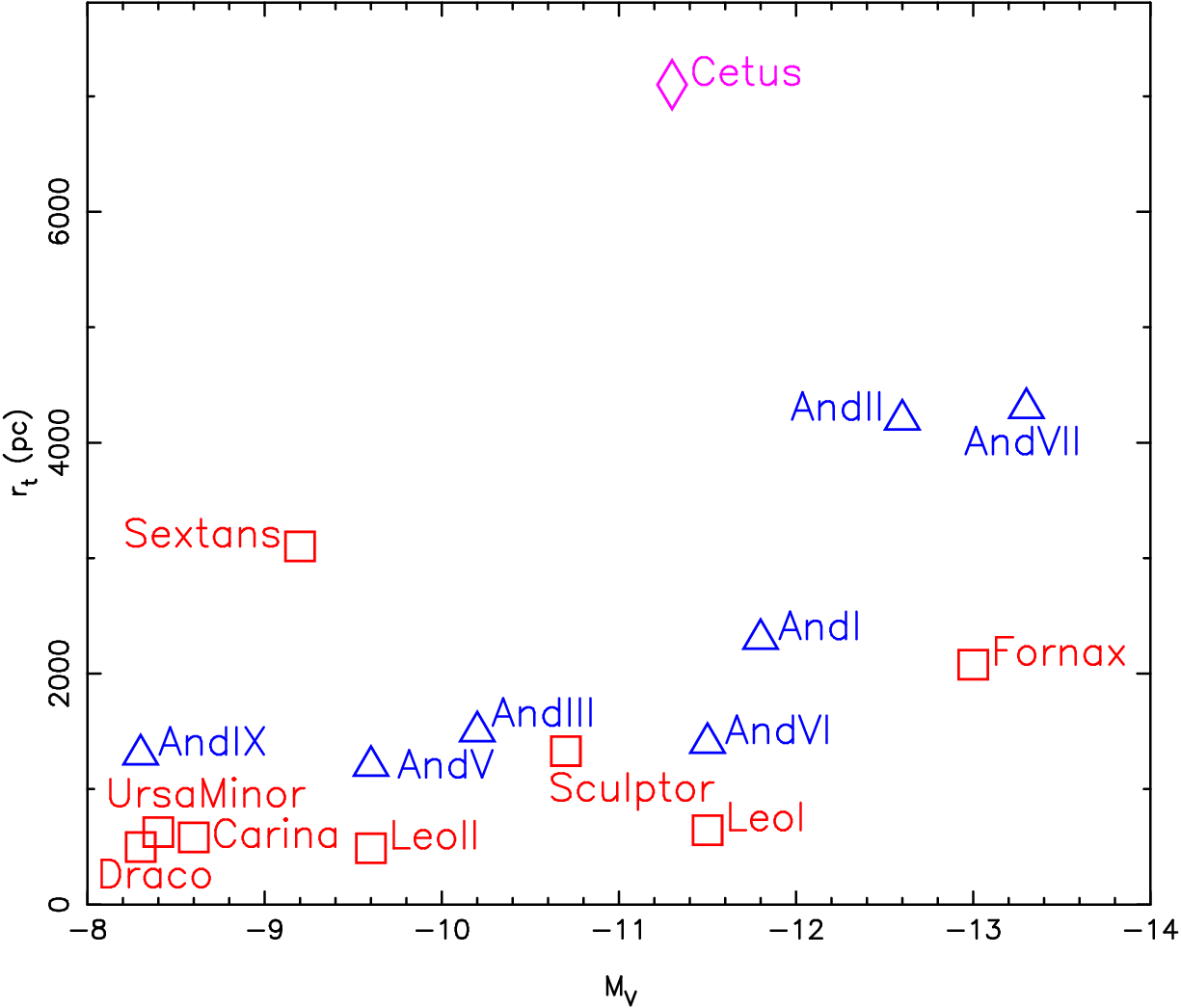}
\caption{Absolute magnitude versus $r_{\frac{1}{2}}$
  (left panel) and $r_t$ (right panel) for the Galactic
  dSphs (red squares), M31 dSphs (blue triangles) and the isolated
  dSph in Cetus (magenta diamond). For a given $M_v$, the M31 dSphs
  have scale-radii that are generally at least twice as large as those
  for the Galactic dSphs.}
\label{m31gal}
\end{center}
\end{figure*}

The M31 sub-group consists of seven dSph galaxies including
Andromeda~IX (\citealt{zucker2004a,chapman2005}), compared to nine
firmly identified dSphs around the Galaxy.  We do not consider
Sagittarius here, however, due to its extreme nature as a highly
disrupted system near to the Galactic disk.

IH95, and subsequent authors, found that King and exponential profiles
generally fit the surface brightness profiles of the Galactic dSphs
very well. For the dSphs analysed here, we find King models are
usually marginally better fits than exponential and Plummer profiles,
which is not unexpected given that King profiles have an additional
parameter. The latter two profiles fit the data equally as well as
each other. Six out of eight of the galaxies looked at by IH95 show an
excess of stars at large radii in comparison to the best-fitting King
model. Of the seven dSphs analysed here, only Andromeda~V shows clear
evidence of this effect. Interpretation of these stars is a
controversial topic in Local Group astronomy. However, in the case of
Andromeda~V, where the stellar excess is a result of a break in the
surface brightness profile, it is likely that this is a manifestation
of tidal effects, as shown by \cite{choi2002}. If this can also be
shown to be the case for the relevant Galactic dSphs, then this might
suggest that the outer regions of the Galactic dSphs are generally
more disrupted than the M31 dSphs. However, only small samples are
being compared and the evidence is indirect at best; Andromeda~I, III
and V all hint, to varying degrees, that these galaxies may be tidally
perturbed. It should also be noted that the photometry on which the
various analyses of the Galactic dSphs are based extends below the
main sequence turn-off, whereas the INT~WFC photometry of M31 dSphs
only samples bright RGB stars.

Figure~\ref{m31gal} shows the absolute magnitude of each dSph plotted
against $r_{\frac{1}{2}}$ (left panel) and $r_t$ (right panel) for
most of the Local Group dSphs. Red squares represent the Galactic
dSphs, blue triangles represent the M31 dSphs, and Cetus is shown as a
magenta diamond. Data are taken from this study, IH95,
\cite{zucker2004a} and \cite{harbeck2005}. The half-light radius of
Andromeda~IX is calculated by integration of the best fit King model
derived by \cite{harbeck2005}, using the distance calculated in
\cite{mcconnachie2005a}. For consistency, we have used the scale-radii
corresponding to the single component fit for Andromeda~II.

Figure~\ref{m31gal} demonstrates that the M31 and Galactic dSphs have
a similar range of $M_V$; the only noticeable difference in this
respect is that there are five Galactic satellites with $M_v > -10$
compared to two for M31.  Although this is a possible indication of 
relative incompleteness of the M31 dSph population, the small numbers
involved mean that a K-S test shows that the two distributions
of $M_V$ are still consistent at the 62\,\% level.

An obvious feature of Figure~\ref{m31gal} is that the values for
$r_{\frac{1}{2}}$ and $r_t$ of the M31 dSphs are much larger than for their 
Galactic counterparts.
For a given $M_V$, the scale-radii of the M31 dSphs are generally twice as 
large than for the Galactic dSphs. Specifically, the mean
and median $r_t$ for the M31 dSphs are 2.0 and 2.3 times larger,
respectively, than for the Galactic dSphs and the mean and median
$r_{\frac{1}{2}}$ are both 3.1 times larger.  The half-light radius,
$r_{\frac{1}{2}}$, tracks $r_c$, $r_e$ and $b$,
and so the same disparity is also seen in these quantities.  As these
differences are observed across the range of $M_V$ presented by the
dSphs, it is unlikely to be an artifact of small number statistics.
Instead, these findings point to notable differences between the formation
and/or evolution of these dSph populations.

A significant difference in the tidal radius, $r_t$, between the two 
populations will,
by default, lead to a difference in the values for $r_{\frac{1}{2}}$. As
demonstrated by the value for Cetus, $r_t$ depends strongly on environment. The
value of $r_t$ derived from a King profile fit is not necessarily the
true $r_t$ for a dSph, but is a useful parameterisation to compare with
simple analytic models. \cite{oh1995} give the following
expressions for the value of $r_t$ for a dSph in the tidal field
induced for a point mass;

\begin{equation}
r_t \simeq \frac{\left(1 - e\right)}{\left(3 + e\right)^{\frac{1}{3}}} a \left(\frac{M}{M_h}\right)^{\frac{1}{3}}
\label{ptmass}
\end{equation}

\noindent and for a logarithmic halo,

\begin{equation}
r_t \simeq \left[\frac{\left(1 - e\right)^2}{\left[\frac{\left(1 + e\right)^2}{2 e}\right] \ln\left[\frac{\left(1 + e\right)}{\left(1 - e\right)}\right] + 1}\right]^{\frac{1}{3}} a \left(\frac{M}{M_h}\right)^{\frac{1}{3}}~.
\end{equation}

\noindent Here, $e$ is the eccentricity of the orbit of the dSph which
has a semi-major axis $a$. $M$ is the total mass of the dSph and $M_h$
is the mass of the host galaxy contained within the current position
of the dSph. The details of these equations are not important for the
current discussion and we include them only to highlight the factors on
which $r_t$ depends. Detailed treatments of $r_t$ also require that
the orbits of the individual stars in the dSph are taken into account,
as stars on radial orbits are preferentially stripped to stars on
circular orbits (\citealt{read2005}).

If we make the plausible assumption that the only fundamental
difference between the M31 and Galactic dSphs is that one group orbits
M31 and the other group orbits the Galaxy, then the differences in
$r_t$ (and the more robust measure $r_{\frac{1}{2}}$) must primarily
arise via some combination of the relative distribution of semi-major axes,
$a$ modulated by the orbital eccentricities $e$, or the dynamical mass
distribution embodied in the parameter $M_h$.  However, it is difficult to 
break the degeneracy of these factors without more detailed modelling. 
For example, if $M_h$ as a function of radius is different between M31 and 
the Galaxy, then so too will the
typical values of $a$. Alternatively, if the characteristic orbits of
the two populations have significantly different values for $e$, then
the fraction of the mass of the host galaxy contained within the
positions of the dSphs could be very different and change as function
of orbital phase, $\theta$. Naively, however, it seems probable that
the difference in the typical values of $r_t$ between the dSphs of M31
and the Galaxy is reflecting a difference in $M_h\left(r,
  \theta\right)$ between these two hosts, and requires detailed
examination.  In this context, it is particularly interesting to note
that \cite{huxor2005} have recently found several extended luminous star
clusters in the halo of M31 with large core and tidal radii, which do not have
any Galactic counterparts.

\subsection{The Local Group population of dwarf spheroidals}
\label{similarities}

\begin{figure}
\begin{center}
\includegraphics[angle=270, width=7.5cm]{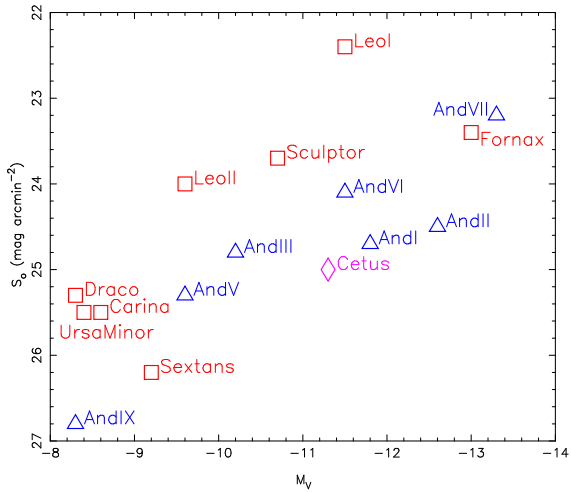}
\caption{The relation between integrated
luminosity and central surface brightness shown by the Local Group
dSphs. For the M31 dSphs, the central surface brightness is that which
has been directly measured from the integrated flux distribution. For
a given $M_V$, the M31 dSphs have systematically fainter central
surface brightnesses, presumably related to their large physical
extent.}
\label{lumsb}
\end{center}
\end{figure}

\begin{figure}
\begin{center}
\includegraphics[angle=270, width=7.5cm]{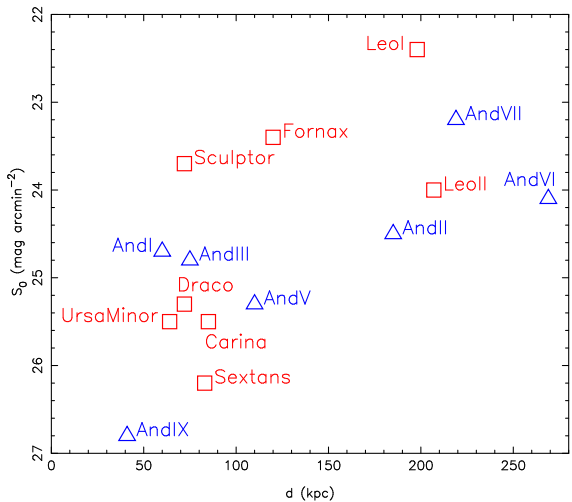}
\caption{Left panel: central surface brightnesses
of the Local Group dSph satellites as a function of distance from the host
galaxy.  For the M31 dSphs, the central surface brightness is that
which has been directly measured from the integrated flux
distribution. The central surface brightnesses of the Local Group dSph
satellites seem to correlate with their current separation from their
host, such that more distant dSphs are brighter. A Spearman rank-order
correlation test indicates that this correlation is significant at the
99\,\% level.}
\label{distcorr}
\end{center}
\end{figure}

The Local Group dwarf galaxies display several correlations between
their physical properties. These include correlations between
luminosity -- halo virial velocity, luminosity -- metallicity,
spin-parameter -- central surface brightness, and central surface
brightness -- luminosity. \cite{dekel2003} propose that all of these
correlations relate to the role of supernovae feedback in dwarf
galaxies, and is an extension of the ideas proposed by
\cite{dekel1986}. The correlation between central surface brightness
and luminosity is shown in Figure~\ref{lumsb}, and is significant at
$> 99.99$\,\% level. While both the M31 and Galactic dSphs show the same
trend, the two populations appear offset from one another such that
the M31 dSphs have systematically lower central surface brightnesses
for a given $M_V$.  This is presumably related to their more extended
nature discussed in the previous section.

Figure~\ref{distcorr} shows the central surface brightness of the
Local Group dSph satellites plotted against their distance from the
host galaxies. There is an apparent correlation between central
surface brightness and current separation, such that brighter dSphs
are further from their host. A Spearman rank-order correlation test
indicates that this correlation is significant at the 99\,\% level. Due to the
correlation between central surface brightness and $M_V$ highlighted
previously, there is a secondary correlation between $M_V$ and separation
which is significant at the 96\,\% level.

This trend has already been noted for the Galactic satellites;
\cite{bellazzini1996} favoured a physical explanation, whereas
\cite{mateo1998a} and \cite{vandenbergh1999} suggested that this
reveals that the Galactic satellite system is incomplete at the faint
end at large distances. Most recently, \cite{willman2004} have
suggested that some Galactic dSphs fainter than 24\,mag\,arcsec$^{-2}$
at a distance greater than 100\,kpc may not have been able to be
detected. However, \cite{mcconnachie2005b} argue that Galactic
satellites as faint as Sextans would have been found within 200\,kpc
of the Galaxy in the 2/3 of the sky that were analysed as part of the
survey that led to the discovery of Sextans (\citealt{irwin1990}). The
detection limits of this survey are discussed in detail in
\cite{irwin1994}; briefly, for satellites located between $\sim
30$\,kpc and $\sim 400$\,kpc from the Galaxy, the detectability of
their resolved image signature is, to first order, $\propto S_o \times
A_o$, where the first term is the central surface brightness and the
second term is their scale-area. $A_o \propto d^{-2}$ and $S_o \propto
d^{\frac{1}{2}}$, implying detectability scales as
$d^{-\frac{3}{2}}$. This implies that: Fornax, Sculptor and Leo I
would be detectable out to 400\,kpc; Leo II would be detectable out to
350\,kpc; Carina and Draco out to 300\,kpc; Sextans and Ursa Minor out
to 250\,kpc; satellites one magnitude fainter than Sextans or Ursa
Minor would be detected out to 200\,kpc. Given this, it is unlikely
that the observed trend for the Galactic dSphs is a result of
selection effects. Further, the M31 dSphs are here shown to display
the same trend as the Galactic dSphs, although different selection
effects apply to this system (\citealt{mcconnachie2005b}). Clearly,
this trend deserves further attention.

It is important to consider the possibility that the correlation
between central surface brightness and current separation may be
spurious, since the observed separation of a dSph from its host galaxy
is a function of orbital phase. The current galactocentric distance of
a satellite from its host may have no physical relevance in terms of
the formation and evolution of that system. To try to quantify this
effect, we have examined the Keplerian case of a satellite orbiting a
point mass in order to determine for what fraction of an orbital
period the instantaneous separation of a satellite from its host is a
reasonable estimate of the time-averaged separation. This is
equivalent to asking what fraction of a population of satellites on
similar orbits will be at a distance from their host which is a
reasonable representation of their mean distance, when viewed at a
random moment in time.

At a given moment, $t$, during an orbit of period $P$, the orbital
phase is given by $\theta = t/P$ and we can iteratively solve Kepler's
equation for the eccentric anomaly, $E$,

\begin{equation}
E = 2 \pi \theta + e \sin{E}~,
\end{equation}

\noindent for an orbit of eccentricity $e$. This allows the
calculation of the instantaneous radius vector in units of the
semi-major axis,

\begin{equation}
r^\prime = 1 - e \cos{E}~.
\end{equation}

\noindent The time-averaged separation, $\left<r^\prime\right>$, of
the satellite is then given by

\begin{equation}
\left<r^\prime\right> = \int_0^1 r^\prime d\theta
\end{equation}

\noindent and the fraction of time, $f_e\left(\eta\right)$, for which
$r^\prime$ is within a certain fraction of $\left<r^\prime\right>$ is

\begin{equation}
f_e\left(\eta\right) = \int_{\theta_1}^{\theta_2} \eta d\theta~.
\end{equation}

\noindent $\eta = r^\prime/\left<r^\prime\right>$ and $\left(\theta_1,
\theta_2\right)$ corresponds to the range of $\theta$ for which
$r^\prime$ is within the required fraction of
$\left<r^\prime\right>$. The subscript indicates that this is for a
specific value of $e$, and the formalism is readily extended to a
range of $e$ since

\begin{equation}
P\left(\eta\right) = \frac{\int f_e\left(\eta\right) P\left(e\right)
de}{\int P\left(e\right) de}~.
\end{equation}

\noindent For simplicity, we set $P\left(e\right)$ to be constant over
some range of $e$.

\begin{table*}
\begin{center}
\begin{tabular*}{0.55\textwidth}{lrcc}
\hline
 && $0.75 \le \eta \le 1.25$ & $0.67 \le \eta \le 1.33$\\
\hline
All values        & $0 \le e \le 1$     & 63\% & 83\%\\
Circularly biased & $0 \le e \le 0.5$   & 83\% & 96\%\\
Radially biased   & $0.5 \le e \le 1$   & 44\% & 70\%\\
\cite{oh1995}     & $0.3 \le e \le 0.7$ & 52\% & 80\%\\
\hline
\end{tabular*}
\caption{The fraction of satellites in a
  population whose current separation from the host is expected to be
  a reasonable estimate of the time-averaged mean distance, indicated
  by the fraction $\eta = r^\prime/\left<r^\prime\right>$, for
  Keplerian orbits with an even spread in $e$ between the limits
  indicated. In all cases, the current separation is within 33\,\% of
  the mean distance most of the time.}
\label{kepler}
\end{center}
\end{table*}

Table~\ref{kepler} illustrates the results for four different
satellite populations, each with a different spread in
$e$. \cite{oh1995} estimated the range of $e$ occupied by the Galactic
satellites to be $0.3 \le e \le 0.7$. Even for the case where orbits
are radially biased, $r^\prime$ is still a reasonable indicator of
$\left<r^\prime\right>$ (ie. within 33\,\%) in $> 2/3$ of the cases. These idealised
calculations suggest that a correlation of $r^\prime$ with central
surface brightness for the Local Group dSph population implies a
similar correlation between $\left<r^\prime\right>$ and central
surface brightness. This then implies that an explanation for
the trends is rooted in physics.

\cite{abadi2005} suggest that, in the current hierarchical paradigm,
the dwarf galaxies observed as satellites today were accreted
relatively recently, and did not form in situ. If this is the case,
then the above correlation is unlikely to be a result of the formation
of the dwarfs but is more likely to be due to their subsequent
evolution. It is difficult to significantly modify the characteristics 
of the central regions of a dwarf galaxy with an external tidal
field, without severely disturbing/disrupting much of the system, since
the inner regions of dSphs are more tightly bound and
shielded than the outer regions (e.g.  \citealt{oh1995}).  The
most significant interactions between the satellites and their hosts
are gravitational.  While ram pressure effects and supernova-driven winds  
undoubtedly contribute to the development of the baryonic density distribution,
it seems likely that gravitational effects play a major role in the 
evolution of the central surface brightness of satellites as they orbit 
and interact with their host.  These effects are directly amenable to study 
by numerical simulations.

\section{Summary}

In this paper, we have derived isopleth maps, surface brightness
profiles, intensity-weighted centres, position angles, ellipticities,
tidal radii, core radii, concentration parameters, exponential scale
lengths, Plummer scale lengths, half-light radii, absolute magnitudes
and central surface brightnesses for the Andromeda~I, II, III, V, VI,
VII and Cetus dSph galaxies. We have shown that the M31 dSph
population shows a large variation in morphology, magnitude and radial
extent. Andromeda~III and V show tentative evidence of tidal
harassment, while the morphology of Andromeda~I clearly identifies it
as a strongly disrupted satellite of M31, suggesting that it may have
extended tidal tails and perhaps only a relatively shallow dark matter
potential. The isolated dSph in Cetus, on the other hand, does not
show any evidence of tidal truncation and has a large radial extent.
These results suggest a wide range of tidal effects is experienced by
Local Group dSphs. Andromeda~II has a clear excess of stars in its
central regions and provides compelling evidence for a
multiple-component system, similar perhaps to Sculptor or Sextans
(\citealt{tolstoy2004,kleyna2004}). Several of the dSphs are found to
be more extended and brighter than previous studies, and in particular
Andromeda~V is shown to have a luminosity consistent with the
metallicity measurements of \cite{armandroff1998} and
\cite{mcconnachie2005a}. Its previous value was consistent
with a lower metallicity derived by \cite{davidge2002}.

In general terms, the M31 dSphs show qualitatively similar structural
characteristics to the Galactic population. However, the M31 dSphs
have much larger scale-radii for a given $M_V$ than the Galactic
dSphs, by a factor of 2 -- 3. We suggest that these differences may
result from differences in the tidal fields experienced by the two
populations, such that the stronger Galactic tidal field truncates its
satellites at smaller average radii in comparison to the M31 dSphs
which evolve in a comparatively weaker tidal field. This implies the
masses of M31 and the Galaxy differ and/or the characteristic orbits
of the satellites are different, and requires detailed modelling. In
this context, it is particularly interesting to note that
\cite{huxor2005} have recently found several luminous extended star clusters in
the halo of M31 with large scale radii, which do not have any Galactic
counterparts. Finally, we highlight a correlation between the central
surface brightnesses and the current separations of the dSphs from
their hosts, which we argue is due to some unknown physical mechanism,
which is most likely dynamical in origin.

\section * {Acknowledgements}
This work is based on observations made with the Isaac Newton
Telescope on the Island of La Palma by the Isaac Newton Group in the
Spanish Observatorio del Roque de los Muchachos of the Institutode
Astrofisica de Canarias. We would like to thank the anonymous referee
for his/her comments, which led to an improvement in the clarity of
this paper. Also, we thank Beth Willman for her feedback on an earlier
draft.

\bibliographystyle{apj}
\bibliography{/pere/home/alan/Papers/references}

\end{document}